\documentstyle[sprocl,epsf]{article}
\def\Journal#1#2#3#4{{#1} {\bf #2}, #3 (#4)}
\def\AP{{\em Ann. Phys.}}

\def\NPB{{\em Nucl. Phys.} B}

\def\PRA{{\em Phys. Rev.} A}

\def\PRD{{\em Phys. Rev.} D}
\def\PREPC{{\em Phys. Rep.} C}

\def\RMP{{\em Rev. Mod. Phys.}}

\newcommand{\be}{\begin{equation}}
\newcommand{\ee}{\end{equation}}
\newcommand{\bea}{\begin{eqnarray}}
\newcommand{\eea}{\end{eqnarray}}

\newcommand{\hf}{{1\over2}}

\newcommand{\nonu}{\nonumber\\}

\begin{document}

\title{RENORMALIZATION GROUP WITH CONDENSATE}

\author{JEAN ALEXANDRE}
\author{VINCENZO BRANCHINA}
\address{Laboratory of Theoretical Physics, Louis Pasteur University\\
3 rue de l'Universit\'e 67087 Strasbourg, Cedex, France\\
alexandr@sbghp4.in2p3.fr, branchina@sbgvax.in2p3.fr}

\author{JANOS POLONYI}

\address{Laboratory of Theoretical Physics, Louis Pasteur University\\
3 rue de l'Universit\'e 67087 Strasbourg, Cedex, France\\
polonyi@fresnel.u-strasbg.fr}
\address{Department of Atomic Physics, L. E\"otv\"os University\\
Puskin u. 5-7 1088 Budapest, Hungary}
\maketitle\abstracts{
The renormalization group is applied to the 
$\phi^4$ model in the symmetry broken phase in order to identify
different scaling regimes. The new scaling laws reflect nonuniversal behavior
at the phase transition. The extension of the analysis to finite
temperature is briefly outlined.  It is mentioned that the coupling constants
can be found in the mixed phase by taking into account the saddle points of
the blocking procedure.}

\section{Introduction}
In the search of the quark confinement mechanism the haaron model has been
proposed\cite{haarons} because it comprises the lesson to be learned from
lattice QCD. The characteristic feature of the model is that the linearly rising
potential between static color charges arises from a simple sine-Gordon type
effective model after a partial resummation of the Haar-measure vertices of the 
path integral. This was a rather puzzling result since it was difficult to accept 
that the leading long range force comes from vertices which are nonrenormalizable 
and irrelevant. But one can easily find the explication of this apparent paradox: 
On the one hand, the renormalizability stands for the relevant or marginal
behavior of the operators in the ultraviolet scaling regime. On the other hand,
the confining forces are observed beyond this scaling regime, where new scaling laws 
arise at the low energies. There is no reason to expect that the relevant operator
set agree for the ultraviolet and the infrared scaling regime. 
Thus one is left with a more general question whether the
set of relevant operators in a model may differ in different scaling regimes
and what consequences such a phenomenon might have. 

When the mass scale is
explicitly given in the lagrangian as in the massive $\phi^4$ model then
the simple perturbation expansion is sufficient to study the different
scaling regimes. There are cases when a partial resummation of the
perturbation expansion is needed to generate the mass gap, like for photons
at finite temperature. In these cases where perturbation expansion applies
there is no transmutation of the degrees of freedom, i.e. one finds
the same particles at every energy and each fixed point is Gaussian. 
When the low energy scaling law is accompanied by the appearance of new particles, 
such as in the two dimensional Gross-Neveu model or in QCD then new relevant 
operators are expected which are responsible for the formation of the new 
composite particles and the low energy fixed point is not Gaussian anymore.
The generic mechanism for the nonperturbative modification of the scaling
law and the generation of new relevant operators is the condensation.
This phenomenon may occur either at the low\cite{sbprd} or at the high 
energy\cite{herve} scaling regimes. Since the mass generation is usually 
achieved in High Energy Physics by the help of the condensates one may find 
similar complication in the unified models, as well, despite their perturbative 
appearance. The present contribution is a brief summary of some results obtained 
in this direction. 

We shall first argue in Section 2 that the appearance of several fixed points and 
scaling regimes is a rule rather than an exception in High Energy Physics.
The powerful Wegner-Haughton form of the renormalization group equation
is introduced in Section 3 as the method to tackle our problem in the 
symmetry broken phase of the $\phi^4$ model. A diverging and a focusing
effect of the renormalization group flow is discussed in Section 4. 
Section 5 is a brief digression into the structure of the mixed phase. The 
generalization of our results to finite temperature is the subject of
Section 6. Finally Section 7 is for the summary.

\section{Multiple Fixed Points\label{mfp}}
The models with intrinsic mass scales possess at least two distinct 
scaling regions, one at the ultraviolet and another one at the infrared side of the
mass scale. The infrared scaling is usually called trivial because it can 
be proven that there is no non-Gaussian relevant or marginal operator.
In fact, for models with finite correlation length the 
fluctuations are exponentially suppressed at large distances and the
evolution of the running coupling constants slows down in the infrared
limit. The manifold of the possible attractive infrared fixed points
is parametrized by the initial values of the relevant 
coupling constants of the ultraviolet scaling regime given at the scale
of the ultraviolet cutoff. When the excitation spectrum has no gap above
the vacuum or there is an instability then the long range interactions might 
be so strong as to change this simple situation. The result is that 
divergences might pile up and drive the trajectory away to another,
unexpected region in the coupling constant space.
These runaway trajectories indicate the existence of a relevant operator 
in the infrared scaling regime.

The presence of several scaling regimes is easily recognizable at the Theory
of Everything. Whatever theory will proven to be that, its renormalized 
trajectory should be traced down in a space which contains all coupling
constants what is used in physics at finite energies.
From the coupling constants of possible composite models at
so far unexplored high energies through the parameters of the Standard Model 
down to the coupling constants in Solid State Physics one 
includes several axes in this space. On the renormalized trajectory depicted 
in fig. 1 one observes the scaling laws characteristic of different
interactions in the energy regime where the trajectory is in the linearizability 
region of a fixed point. Note that the trajectory may be influenced by the 
environment and bifurcates into different thermodynamical 
phases in the IR regime in different laboratories\cite{cargese}.

\begin{figure}\label{rgtoe}
	\epsfxsize=12cm
	\epsfysize=5cm
	\centerline{\epsffile{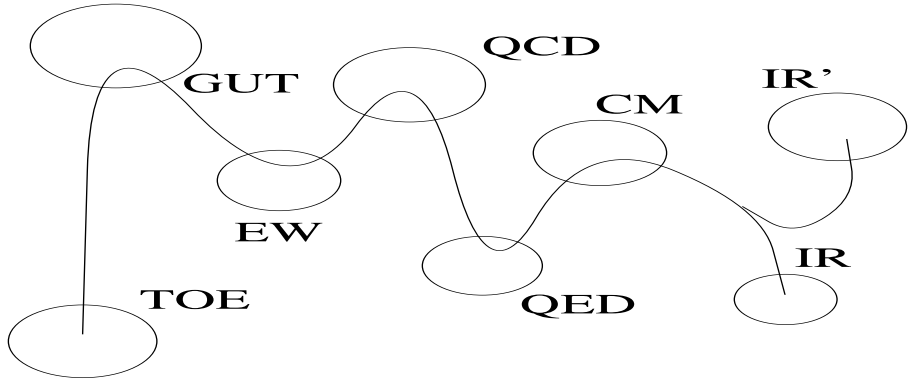}}
\caption{The renormalized trajectory of the Theory of Everything (TOE)
starts at the (supposed) ultraviolet fixed point at $k=\Lambda$, 
it passes by the fixed points of the 
Grand Unified Models (GUT), the unified Electro-Weak theory (EW), 
the strong interactions (QCD), the electromagnetic interactions (QED) 
certain fixed points of the Solid State and Condensed Matter Physics (CM) 
and finally approaches the ultimate IR fixed point, $k\to0$. The
circles denote the domains of linearizability, the asymptotic
scaling regions.}
\end{figure}

It is mathematically certainly correct to say that
the renormalized trajectory, the set of the physical "constants",
is given by the initial condition at the TOE. In this respect the
High Energy physicist seeks the few ultimate constants of Nature.
But the same algebra of observables is classified at each scaling region
according to the actual scaling laws and there is no guarantee that
the set of relevant and marginal operators is found to be always the same.
The renormalizable (relevant or marginal) operators of a scaling regime are 
usually nonrenormalizable (irrelevant) at the higher energy fixed points
such as the quark-gluon QCD vertex appears as a nonrenormalizable 
one in a composite model for the quarks and gluons. Consider now a 
coupling constant, denoted by $g_{n-r}(k)$ what is nonrenormalizable at high energy 
and becomes relevant at a lower energy scaling regime. Then this 
coupling constant undergoes a suppression at the high energy scaling regime.
How can it
became a relevant coupling constant at the low energy scaling regime? What is its role 
during the evolution? According to the usual scenario the initial value of the 
irrelevant coupling constant, $g_{n-r}(\Lambda)$, modifies the theory in an overall scale 
and the relation between observables of the same dimension is given by the initial value
of the renormalizable coupling constants only. This is expressed by the
condition for the beta function for any coupling constant $g$ as
\be
\lim_{\Lambda/k\to\infty}{\partial\beta_g(k)\over\partial g_{n-r}(\Lambda)}=
\lim_{\Lambda/k\to\infty}k{\partial^2g(k)\over\partial k\partial g_{n-r}(\Lambda)}=0,
\label{univ}
\ee
where the coupling constants are made dimensionless by the help of the
running cutoff, $k$.

One can imagine the following two, opposite possibilities as far 
as the coupling constant $g_{n-r}$ is concerned:

\begin{enumerate}
\item {\em Divergence:} The universality, eq. (\ref{univ}) is violated
because the trajectories with slightly differing initial conditions
for $g_{n-r}$ diverge from each other. The initial value of $g_{n-r}$ must 
then be specified at the TOE and it becomes an independent free parameter. 

\item {\it Focusing:} The dimensionless quantities at low energy
are (at least locally) independent of 
the initial values of the coupling constants of the TOE and are determined by 
one of the lower energy fixed points. This is a strong version of the
universality because the coupling constants at low energy 
are independent of the initial values of the renormalizable coupling constants.
\end{enumerate}

The existence of different scaling regimes may lead to serious problems 
in indentifying the
important parameters of the theory. In both cases mentioned above the
determination of the physical content of the theory at different energies
in terms of the initial values of the relevant coupling constants of the
TOE is, though being mathematically possible, unfeasible by means of
measurements with small but finite errors. 

\begin{table}\label{qedcc}
\begin{center}
\begin{tabular}{@{}*{4}{|l|l||l|}}
\hline    
U.V.&I.R.&Fig. \ref{qedccp}&example\cr 
\hline    
\hline
relevant&relevant&(a)&$m_e\bar\psi_e\psi_e$\cr
relevant&irrelevant&(b)&$m_\mu\bar\psi_\mu\psi_\mu$\cr
irrelevant&relevant&(c)&$G_4(\bar\psi_e\psi_e)^2$\cr
irrelevant&irrelevant&(d)&$G_6(\bar\psi_e\psi_e)^3$\cr
\hline
\end{tabular}
\end{center}
\caption{The four classes of the coupling constants in QED.}
\end{table}

\begin{figure}
\begin{minipage}{12cm}
\begin{minipage}{5.5cm}
	\epsfxsize=5cm
	\epsfysize=3cm
	\centerline{\epsffile{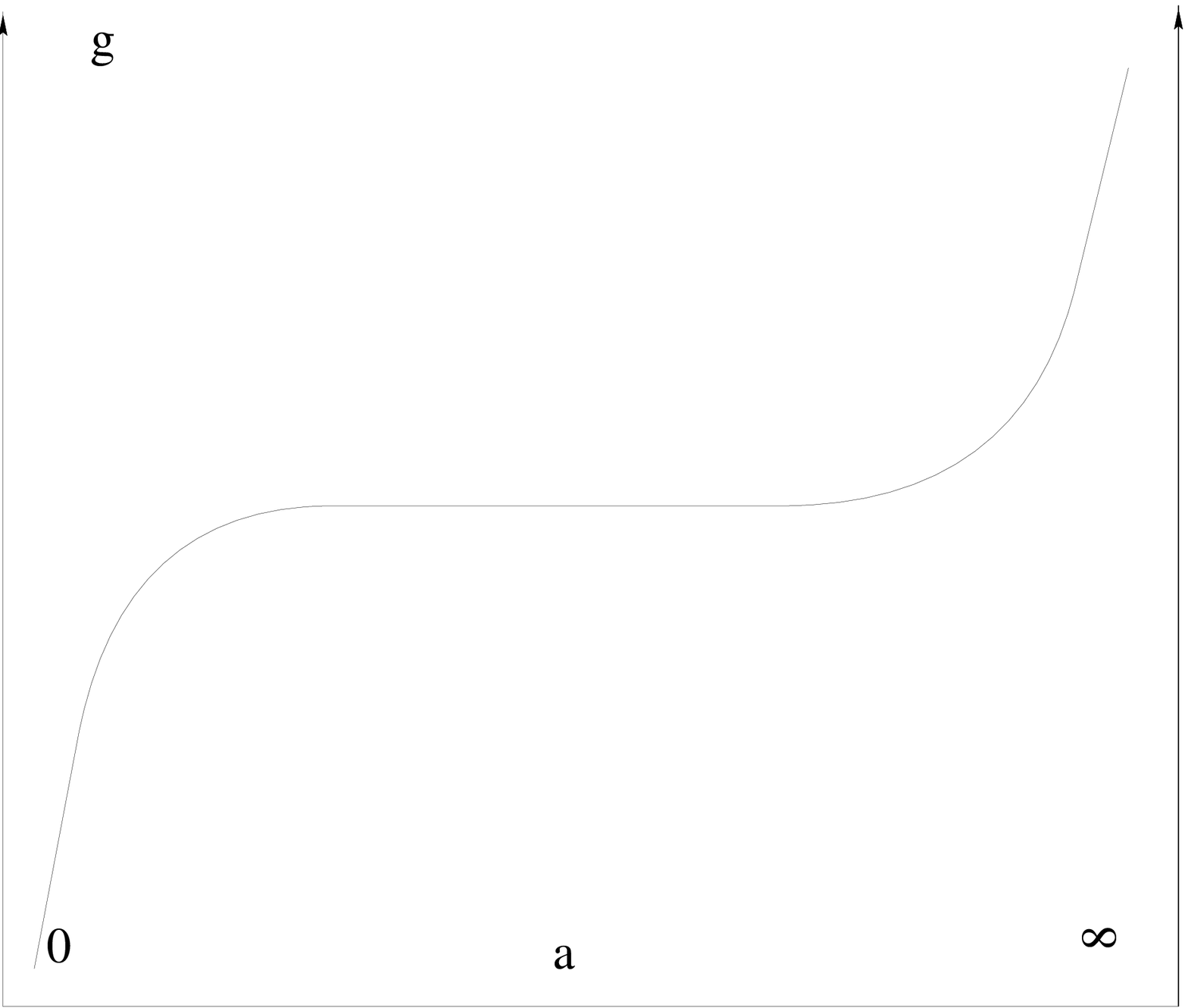}}
\centerline{(a)}
\end{minipage}
\hfill
\begin{minipage}{5.5cm}
	\epsfxsize=5cm
	\epsfysize=3cm
	\centerline{\epsffile{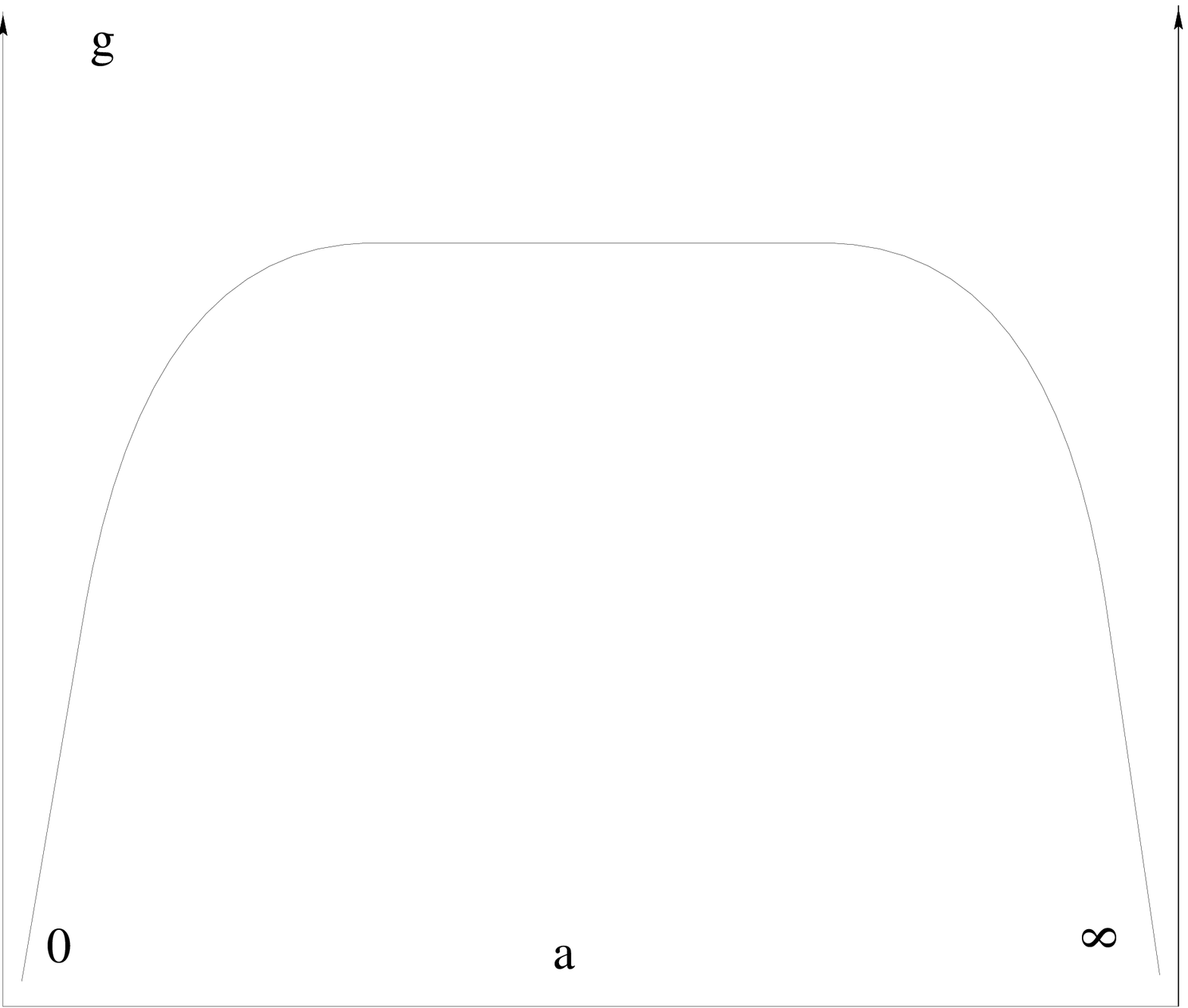}}
\centerline{(b)}
\end{minipage}
\end{minipage}
\begin{minipage}{12cm}
\begin{minipage}{5.5cm}
	\epsfxsize=5cm
	\epsfysize=3cm
	\centerline{\epsffile{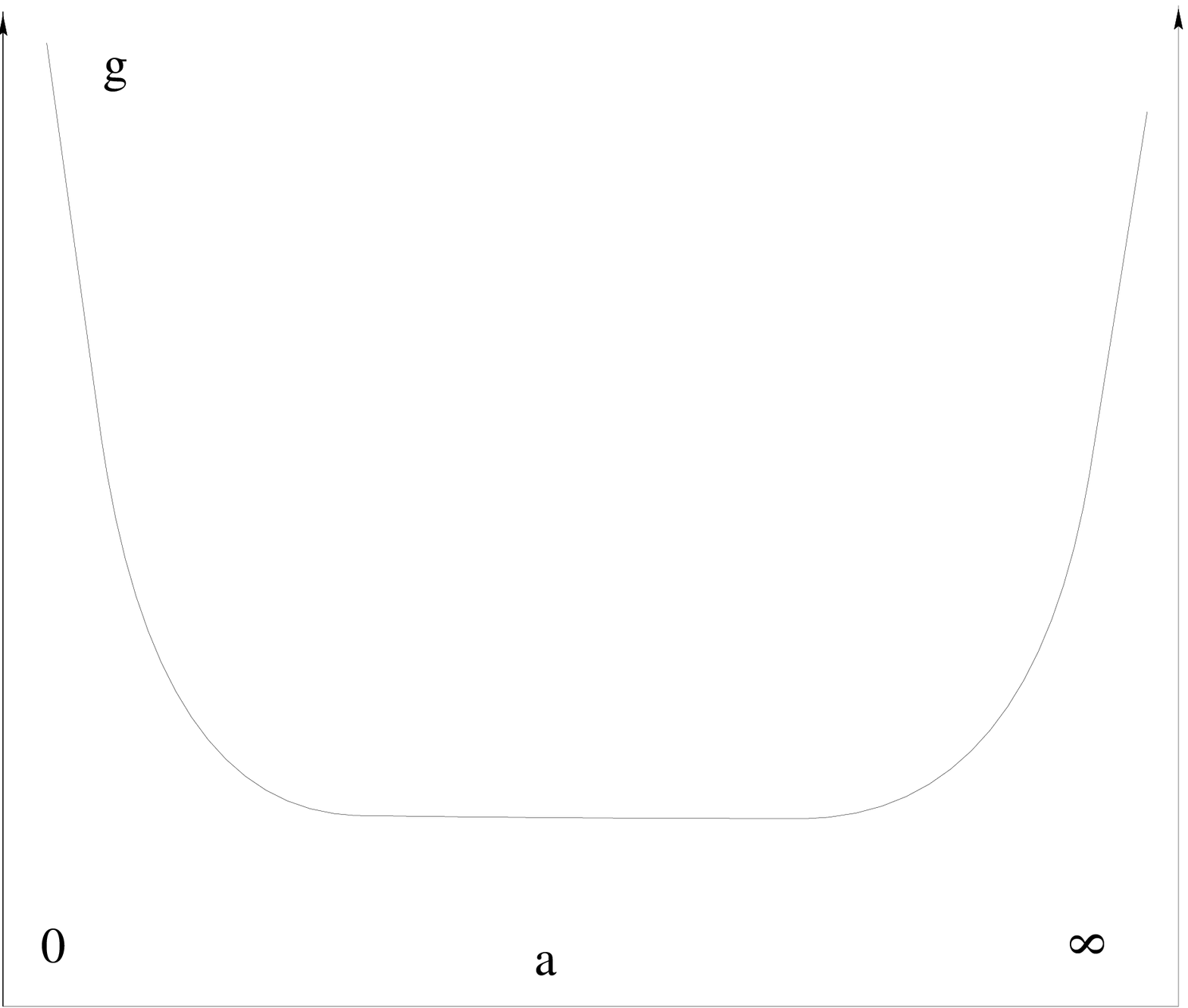}}
\centerline{(c)}
\end{minipage}
\hfill
\begin{minipage}{5.5cm}
	\epsfxsize=5cm
	\epsfysize=3cm
	\centerline{\epsffile{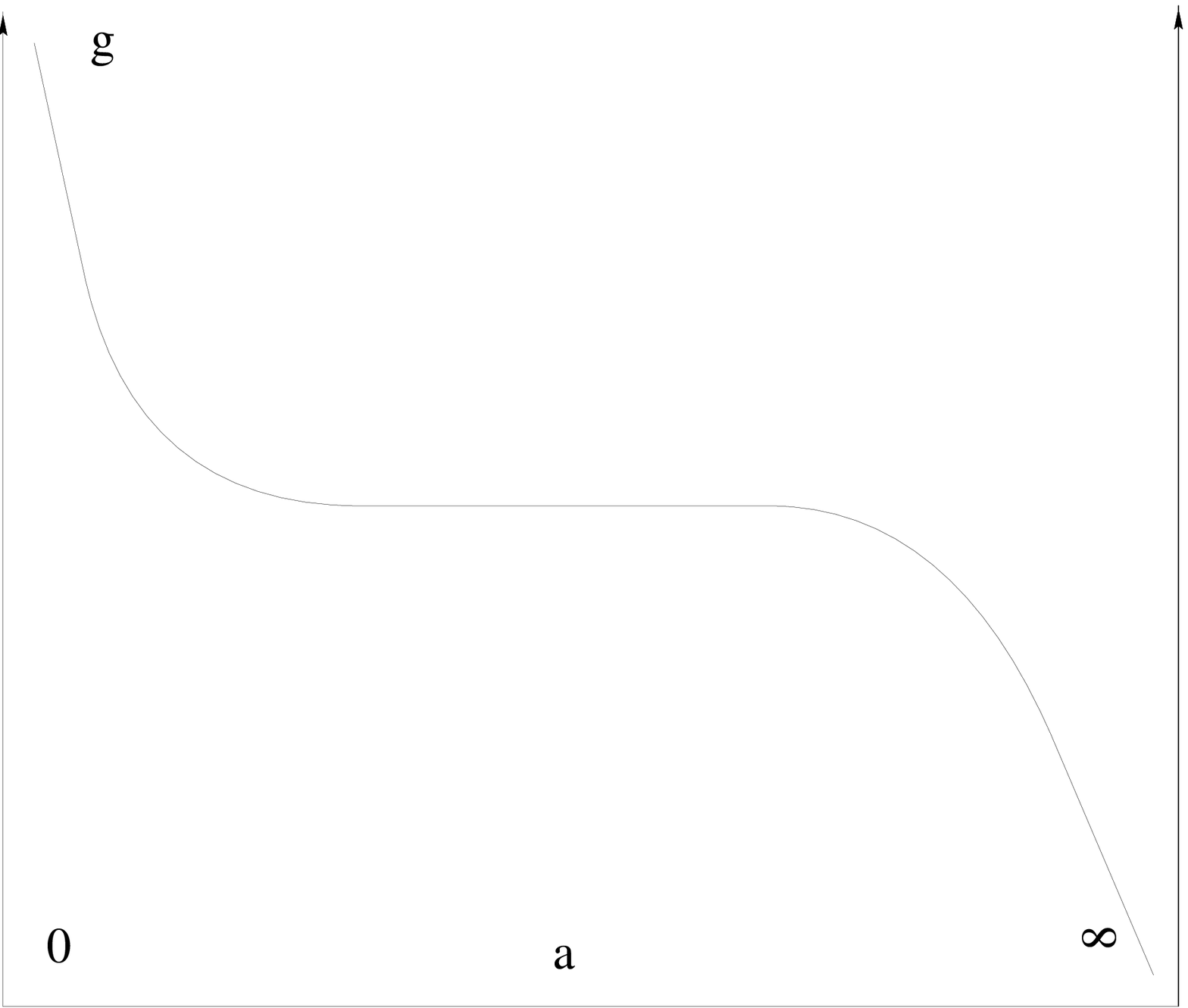}}
\centerline{(d)}
\end{minipage}
\end{minipage}
\label{qedccp}
\caption{The qualitative dependence of the running coupling constants
in the function of the cutoff, $a=2\pi/\Lambda$. The UV and the IR scaling
regimes are shown. The coupling constant is supposed to be
constant in between for simplicity. See Table 1 for the details.}
\end{figure}

It is instructive to consider a simpler model
with two scaling regimes whose generic scaling patterns
are listed in table 1. Consider the theory of photons, electrons and nuclei
in the presence of a chemical potential for the baryon number. For certain values
of the chemical potential the ground state is a solid state lattice. We
identify two scaling regimes, the ultraviolet one characteristic of QED in
the trivial, homogeneous vacuum and an infrared one which is governed by the
lattice effects. The electron mass, $m_e$, is a relevant parameter in each scaling regimes
since it is a renormalizable coupling constant of QED and appears in the
equations of Solid State Physics. The muon mass, $m_\mu$, is as renormalizable parameter
at high energy as $m_e$ but drops out from the physics of the
solids because the processes (without neutrinos) at the energies of eV 
the muon contributions are always dominated by the contributions from the electron.
The six-electron vertex is irrelevant everywhere. The four electron vertex
with the coupling constant $G_4$ is the most interesting interaction. 
It is a nonrenormalizable, irrelevant vertex of QED. But it becomes
relevant at low energies since it is the effective vertex which
is generated by the attractive force between the Cooper-pairs and 
drives the transition into the superconducting state. In the simple
perturbative treatment of QED $G_4$ is suppressed in the ultraviolet
scaling regime so its high energy initial value in the QED lagrangian
is set to zero. It was only after the experimental discovery of the
superconducting phase and its explication by the BCS ground state
that the importance of $G_4$ at lower energies was demonstrated by means
of the partial resummation of the perturbation expansion. This is an example where
the renormalization group was used to find a coupling constant what appears to be 
unnecessary at high energies but nevertheless is important at low energy. Such
parameters will be called hidden coupling constants after their undetectable small values 
at intermediate energies. 

The divergence, case 1 mentioned above, requires that the 
suppression of $g_{n-r}$ in the ultraviolet regime is weaker than the amplification during 
the low energy scaling. If the fixed point of the infrared scaling has
a certain region of the ultraviolet coupling constants in its attractive
zone then focusing, the case 2, is realized. 
We present here a study of the four dimensional $\phi^4$ model
in search for the manifestations of these phenomena.

\section{Wegner-Haughton Equations\label{swh}}
In order to study the role of nonrenormalizable operators on the evolution
one has to be able to handle the mixing of a large number of operators during
the renormalization group transformation. This is achieved in an elegant
manner by the Wegner-Haughton equation \cite{weho}. This is a functional 
differential equation describing the evolution of the renormalized, blocked
bare action under the change of the cutoff. According to the usual strategy
the running coupling constants are identified with the bare ones and the
observational scale with the cutoff so the evolution of the bare coupling
constants qualitatively reproduces the trajectory for the running coupling
constants. We shall be satisfied here to indicate the derivation of the
leading order equation in the gradient expansion\cite{sbap}, the preliminary
results indicate that the higher orders do not change our conclusions.

Let us write the action corresponding to the ultraviolet cutoff $k$ as
\be
S_k[\phi]=\int d^dx\left[\hf(\partial_\mu\phi(x))^2+U_k(\phi(x))\right].
\label{lagrlo}
\ee
According to the usual Wilson-Kadanoff blocking procedure\cite{wilsrg}
we write
\be
e^{-{1\over\hbar}S_{k'}[\phi]}=\int D[\phi']e^{-{1\over\hbar}S_k[\phi+\phi']}
\label{brg}
\ee
where $k'<k$ in the Euclidean space-time. The Fourier transform of the field 
variable $\phi$ and $\phi'$ are nonvanishing for $p<k'$ and $k'<p<k$,
respectively. The right hand side is evaluated by means of the loop expansion,
\be
S_{k'}[\phi]=S_k[\phi+\phi'_0]+{\hbar\over2}\mbox{tr}\log\delta^2S+O(\hbar^2),\label{elim}
\ee
where
\be
\delta^2S(x,y)={\delta^2S_k[\phi+\phi'_0]\over\delta\phi'(x)\delta\phi'(y)},
\ee
and the saddle point, $\phi'_0$, is defined by the extremum condition
\be
{\delta S_k[\phi+\phi'_0]\over\delta\phi'(p)}=0,
\ee
in which the infrared background field, $\phi(x)$, is held fixed.
Eq. (\ref{elim}) is the generalization of the Wegner-Haughton equation for
condensates. One can prove that the saddle point is trivial, $\phi'_0=0$, so long as the 
matrix $\delta^2S(x,y)$ is invertible and the infrared background field is
homogeneous, $\phi(x)=\Phi$. What is remarkable in this equation is that each
successive loop integral brings a suppression factor 
\be
{k^d-k'^d\over k'^d}=O\left({k-k'\over k'}\right)
\ee
due to the integration volume in the momentum space. By choosing an infinitesimal
fraction of the degrees of freedom to be eliminated in a step we find a new
small parameter, $\delta k/k'=k-k'/k'$, suppressing the higher loop contributions
in the blocking relation.

In the leading order of the gradient expansion, the so called local
potential approximation, the only function
characterizing the action is the local potential, $U_k(\phi)$, so we choose
the infrared background field homogeneous, $\phi(x)=\Phi$, and obtain from (\ref{elim})
\be
U_{k-\delta k}(\Phi)=U_k(\Phi)+\hf\int{d^dp\over(2\pi)^d}\log\left[p^2+U_k''(\Phi)\right]
+O(\delta k^2),\label{fdwh}
\ee
where we introduced the notation
\be
U_k''(\Phi)={\partial^2U_k(\Phi)\over\partial\Phi^2}.
\ee
In the limit $\delta k\to0$ one ends up with the differential equation
\be
k{\partial\over\partial k}U_k(\Phi)=-{\Omega_dk^d\over2(2\pi)^d}
\log\left[k^2+U_k''(\Phi)\right]\label{rgde}
\ee
with
\be
\Omega_d={\Gamma\left({d\over2}\right)\over2^d\pi^{\pi/2}},
\ee
the projection\cite{haha} of the functional equation (\ref{elim}) onto the functional form
(\ref{lagrlo}). It is instructive to expand this equation in
$U_k''(\Phi)-m^2_k$, where 
\be
m^2_k=U''(0),
\ee
when we find
\be
k{\partial\over\partial k}U_k(\Phi)=-{\Omega_dk^d\over2(2\pi)^d}\left[
\log(k^2+m^2_k)+\sum_{n=1}^\infty{1\over n}
\left({U_k''(\Phi)-m^2_k\over k^2+m^2_k}\right)^n\right].
\ee
One recovers here the usual one loop resummation of the effective 
potential\cite{cowe} 
except that the loop momentum is now restricted into the subspace
of the modes to be eliminated. Note that the derivation of eq. (\ref{rgde}) shows 
that the restoring force for the fluctuations into the equilibrium is 
proportional to the argument of the logarithm function. Thus when
\be
k^2+U_k''(\Phi)=0
\ee
then nontrivial saddle point should be used. The original equation, (\ref{elim}),
remains always valid because the action is bounded from below.

We define the coupling constant and the beta function to the $n$-th order 
vertex for the vacuum $\phi(x)=\Phi$ as
\bea
g_n(k)&=&{\partial^n\over\partial\Phi^n}U_k(\Phi),\\
\beta_n&=&k{\partial\over\partial k}g_n(k)
={\partial^n\over\partial\Phi^n}k{\partial\over\partial k}U_k(\Phi),\nonumber
\eea
where in the last equation we assumed the analycity of the potential
in $k$ and $\Phi$, what holds except at the singular points. By taking the 
successive derivatives of the renormalization 
group equation (\ref{rgde}) one obtains
\be
\beta_n=-{\Omega_dk^d\over2(2\pi)^d}{\cal P}_n(G_2,\cdots,G_{n+2}),
\ee
where 
\be
G_n={g_n\over k^2+g_2}
\ee
and
\be
{\cal P}_n={\partial^n\over\partial\Phi^n}\log\left[k^2+U_k''(\Phi)\right]
\ee
is a polynom of order $n/2$ in the variables $G_j$, $j=2,\cdots,n+2$,
\bea\label{pebfv}
\beta_1&=&G_3,\nonu
\beta_2&=&G_4-G_3^2,\nonu
\beta_3&=&G_5-3G_3G_4+2G_3^3,\nonu
\beta_4&=&G_6-4G_5G_3-3G_4^2+12G_3^2G_4-6G_3^4,\\
\beta_5&=&G_7-5G_6G_3-4G_5G_4+20G_3^2G_5-6G_4^2G_5+30G_4^2G_3\nonu
&&-60G_3^3G_4+24G_3^5,
\nonumber
\eea
etc. It is interesting to verify that ${\cal P}_n$ contains the integrand 
of all one loop graphs what contribute to $\beta_n$.

We change now to dimensionless parameters, 
\be
k\to{k\over\Lambda},~~U\to k^dU,~~\phi\to k^{{d\over2}-1}\phi,~~
g_n\to k^{n(1-{d\over2})-d}g_n,
\ee
and
\be
G_n\longrightarrow{g_n\over1+g_2},
\ee
what will be used below. 

One can distinguish an ultraviolet and an infrared scaling regime, for 
$k^2>>|m^2_k|$ and for $k^2<<|m^2_k|$, respectively. In the former one recovers
the usual renormalization group coefficient functions used in studying
the asymptotic scaling. The latter is trivial for $m^2_0>0$ when
the factor $k^d$ suppresses the evolution at the infrared fixed point.
The infrared scaling law is presumable trivial in the case $m^2(0)=0$ 
because the vacuum of this model is supposed to be\cite{cowe} at $\Phi\not=0$ where 
the fluctuations become massive again. 

The symmetry broken phase\footnote{It could naively be identified with $m^2(0)<0$ but 
we should not forget the Maxwell construction what sets $m^2(0)=0$.} is 
characterized by the condition that there is a scale, $k_{cr}(\Phi)$, where the 
spinodial instability occurs and the restoring force of the fluctuations is vanishing,
\be
k_{cr}^2(\Phi)=-U_{k_{cr}}''(\Phi)
\ee
in a region around $\Phi=0$. We shall call the line $k_{cr}^2(\Phi)$ on the plane
$(\Phi,k^2)$ critical. The renormalized trajectory has discontinuous
derivatives along the curve\cite{hahad} and the system is in a mixed 
phase for $k^2<k_{cr}^2(\Phi)$. 
The tree level instability induces nontrivial saddle points
for the blocking (\ref{elim}) and eq. (\ref{rgde}), being based on the
vanishing of the saddle point, is no longer valid. As the critical curve is 
approached in decreasing $k$ during the blocking the denominator $1+g_2(k)$ 
becomes small in the beta functions and new scaling laws develop as a precursor 
of the mixed phase. There is no reason to expect that these new 
scaling laws should be trivial. We shall study the renormalization group flow 
in the vicinity and below the critical line.

One can nowadays experience a renaissance of the infinitesimal renormalization 
group methods. The so called exact renormalization group\cite{exact} is
similar in spirit to the Wegner-Haughton equation but it follows the evolution
of the generator functional for the connected or the 1PI vertices. The advantage 
of this method is that it produces directly the particle physics motivated running 
coupling constants which are based on the scattering amplitudes. The bare
renormalization group, (\ref{brg}), yields simpler expressions for the evolution
of the bare coupling constants of the action. We found this latter method more
attractive since, as mentioned at the beginning of this section,
the evolution of the bare and the running, renormalized coupling constants 
is qualitatively similar. 

Another distinguishing feature of the blocking procedure employed in this work is
that the momentum space cutoff is imposed in a sharp manner. The smooth cutoff
is believed to be superior and its use is more widespread. We shall argue below
that the problems with the sharp cutoff are not incurable with careful
methods and that actually no smooth cutoff is applicable for models 
with instabilities.  The sharp cutoff induces 
diffraction integrals during the blocking what represent
oscillating forces at large distances. The oscillations cast doubt
on the physical significance of the running parameters of the blocked
action and were blamed for the occurrence of the infrared singularities in the
renormalized trajectory. Attempts to eliminate the oscillating
part from the blocked action lead to the introduction of a smooth cutoff.
Let us deal with the cases where the singularity occurs at $k_{cr}^2=0$ and 
$k_{cr}^2>0$ separately and consider a physical quantity obtained in the framework
of the loop expansion,
\be
P(\epsilon,k)=\sum_{n=1}^\infty\hbar^nI_n(\epsilon,k),
\ee
where $I_n(\epsilon,k)$ stands for the $n$-th order loop integral with the
range of integration $\epsilon\le|p_j|\le k$, $j=1,\cdots,n$. In the case
of the infrared unstable theories it is necessary to introduce the infrared cutoff, 
$\epsilon$, what is removed after the computation is completed. The blocking
transformation with sharp cutoff can be used to obtain the right hand side of
the differential equation
\be
{\partial P\over\partial\epsilon}=B(\epsilon).
\ee
The integration of this equation yields the quantity sought when $\epsilon\to0$.
So long as the nontrivial vacuum, e.g. the condensate, what shields the infrared 
divergences is properly incorporated in the computation the thermodynamical 
limit is well defined and the $\epsilon$-dependence is continuous at $\epsilon=0$. 
Moreover any infrared cutoff should yield the same thermodynamical limit.
When the singularity occurs at finite scale then the quantity $P(0,k)$ can
be thought as it had been obtained in the effective theory with a sharp ultraviolet
cutoff at $\Lambda=k$. The bare parameters of this effective theory have singular
cutoff dependence at $k=k_{cr}$. One might argue that this singularity is an
artifact of the cutoff employed because the observables in the effective
theory, being renormalization group invariant, show no singular $k$-dependence.
But we see no conceptual problem with singular, i.e. nondifferentiable
renormalized trajectories if the singularity corresponds to a real physical effect,
i.e. some instability what shows up at a well defined, sharp
value of the momentum. One should simply make sure that 
the physical effects behind the singularity have properly been accounted for
during the solution of the effective theory or the continuation of the 
blocking procedure.

The argument prohibiting the application of any smooth cutoff in theories with
condensate or other instabilities is the following. In order to show the
exactness of the infinitesimal renormalization schemes one has to assume 
first the validity of the loop expansion. In fact, the counting of the
power of $\delta k$ is done in the loop expansion where one has to integrate 
around the saddle point. Without the proper choice of the saddle point the 
formal steps in arguing about the exactness are not valid. 
The point is that the subsequent elimination of the
degrees of freedom modifies the integrand for the unstable mode. Due to the 
factor $\hbar^{-1}$ coming from the nontrivial saddle point the loop corrections
of the stable modes what are computed after the elimination of the unstable mode
yields the contribution $O(\hbar^0)$. Thus all stable modes should completely be
eliminated before one arrives at the unstable sector of the theory.
The models with condensate require a blocking method where the stable modes 
are eliminated completely before arriving at the instability and
the tree level structure of the saddle point expansion must be retained.
Finally we note that this type of cutoff poses no problem in
going to higher orders in the gradient expansion.

\section{Zooming into a fixed point}
Our goal is to follow the evolution of the blocked action from the initial condition
set at $k=\Lambda=1$ towards the infrared regime. We shall consider the one component 
four dimensional 
$\phi^4$ model in the symmetry broken phase. Thus the integration of the 
differential equation (\ref{rgde}) in $k$ runs into a singularity for
$\Phi=0$ at $k=k_{cr}(0)>0$. When the evolution in the outer, stable
region is approximated by the tree level expression, $U_k(\Phi)=U_\Lambda(\Phi)$,
then the singular line, $k_{cr}^2(\Phi)$, is an upside down parabola, 
$k_{cr}^2=-m^2_\Lambda-g_4(\Lambda)\Phi^2/2$, on the plane $(\Phi,k^2)$. But this is an 
inconsistent approximation for the tree level saddle point structure is nontrivial 
in the unstable region for $k<k_{cr}(\Phi)$. Since the input for the elimination 
of a mode at $k$ 
is the potential $U_{k+\delta k}(\Phi)$ for $-\infty<\Phi<\infty$ the result
of the naive blocking what does not take into account the nontrivial saddle point
structure is built on the wrong potential and is incomplete by a term 
$O(\hbar)$ even in the outer, stable region. We shall avoid the problem of the
singularity by assuming that the potential $U_k(\Phi)$ is analytical inside and
outside of the unstable region and its only nonanalytical behavior is confined
on the curve $k_{cr}^2(\Phi)$. This assumption allows us to study the evolution
equation locally in $\Phi$, i.e. by expanding the potential as
\be
U_k(\Phi)=\sum_{n=0}^N{1\over n!}g_n(\Phi_0)(\Phi-\Phi_0)^n
\ee
and following the evolution of the coupling constants $g_n(\Phi_0)$ what
obeys the loop expansion with the trivial saddle point in the outer, stable region. 

We note that the Taylor
expansion of the potential motivated here by the decoupling of the stable and the
unstable region proved to be necessary in the numerical integration of eq.
(\ref{rgde}). The numerical integration of this equation by simply
discretizing the variables $\Phi$ and $k$ becomes highly unstable at 
$k^2\approx k_{cr}^2(0)$ because the logarithm function amplifies the numerical 
errors in computing $U_k''(\Phi)$. The smoothing or interpolating techniques we
tried were unable to stabilize the solution. Thus we finally integrated numerically 
the coupled differential equations imposed at $\Phi_0=0$,
\be
k{\partial g_n\over\partial k}=-(n-4)g_n-{k^4\over16\pi^2}{\cal P}_n(G_2,\cdots,G_{n+2}),
\ee
for $n=1,\cdots,N$.  

We can present here only some preliminary numerical results, the detailed account
will be given elsewhere\cite{jeani}. They suggest the existence of 
two distinct scaling regimes, an ultraviolet, $k^2>>k_{cr}^2(0)$, and another 
one in the vicinity of the singular line. Though the singular line is not a fixed point 
since $k\not=0$ or $\infty$ nevertheless the singularities suggest to parametrize the 
renormalized trajectory by 
\be
\tilde k^2=\cases{k^2-k_{cr}^2(0)&if $k^2>k_{cr}^2(0)$,\cr 0&if $k^2<k_{cr}^2(0)$.}
\ee
This parametrization possess scaling region at at $\tilde k=0$ and $\infty$.
Furthermore we shall argue below that the potential is renormalization group invariant
in the interior, unstable region. In this manner
the whole unstable region represents a single fixed point. Note that though the
critical point belongs to finite scale, $k=k_{cr}(0)\not=0$, the singularities
presented below require a dense enough spectrum for the momentum operator, the 
execution of the thermodynamical limit.

The result of the numerical integration indicates an attractive fixed point
at $k=k_{cr}(0)$  for $N=10$, i.e. by truncating the potential at
$O(\Phi^{20})$.  The value of $\delta k$ was 
adjusted during the integration in the range $10^{-18}<\delta k<10^{-2}$. We
compared the third and the fourth order Runge-Kutta approximation for the
coupling constants and $\delta k$ was chosen to keep the relative local
error on them less than $10^{-15}$.  After the system 
leaves the ultraviolet scaling regime the higher order coupling constants 
undergo oscillations with large amplitude and a new scaling law is found
in the vicinity of the instability. For example $g_{20}$ reaches
the range of $10^{-12}$ for $k\approx k_{cr}(0)$ after 
having gone through the peaks at $10^{23}$.
The fixed point at $\tilde k=0$ corresponds to the potential
\be
U_{k_{cr}(0)}(\Phi)=-\hf k_{cr}^2(0)\Phi^2.
\ee
The infrared fixed point is trivial and its attractive zone seemed 
to extend over the whole symmetry broken phase of the renormalizable $\phi^4$ model. 
The approach to the fixed point is such that $G_n\to0$, $n>2$ as $k^2\to k_{cr}^2(0)$. 
This is consistent with the observation 
that the fourth equation of (\ref{pebfv}) excludes any finite, 
nonzero value for $g_4$ at the critical point what belongs to the trivial root
of ${\cal P}_n$, $n>2$.
Thus the scaling at this critical point shows strong universality, its result does not 
depend even on the renormalizable coupling constants. The symmetry broken $\phi^4$ model
when its potential is truncated at $O(\phi^{20})$ realizes an example 
of the case 2 mentioned in Section \ref{mfp}. 

When the potential is truncated at $O(\Phi^{22})$ or at higher order the qualitative
behavior of the solution is different. At a certain point
the trajectory with $N\ge11$ suddenly turns away from the solution with $N\le10$ and
the coupling constants start to diverge violently towards $-\infty$. The qualitative
behavior seen on fig. 3 remains the same except the singularity becomes 
stronger for $N>11$. The sudden departure of the
trajectories with $N=10$ and 11 can be traced back to the contribution of
$g_{22}$ to $\beta_{20}$ which happens to be exceedingly large at $k\approx k_{cr}(0)$
and starts to push the lower order coupling constants one after the other
towards $-\infty$.
The integration of such a singular curve requires extreme numerical accuracy.
The quadruple precision was used in the codes and the relative
precision of the beta functions was kept below $10^{-11}$. The values of 
$\delta k$ were between $10^{-17}$ and $10^{-13}$ at $k\approx k_{cr}(0)$. 

\begin{figure}\label{evol11}
\begin{minipage}{10cm}
	\epsfxsize=6cm
	\epsfysize=3cm
	\centerline{\epsffile{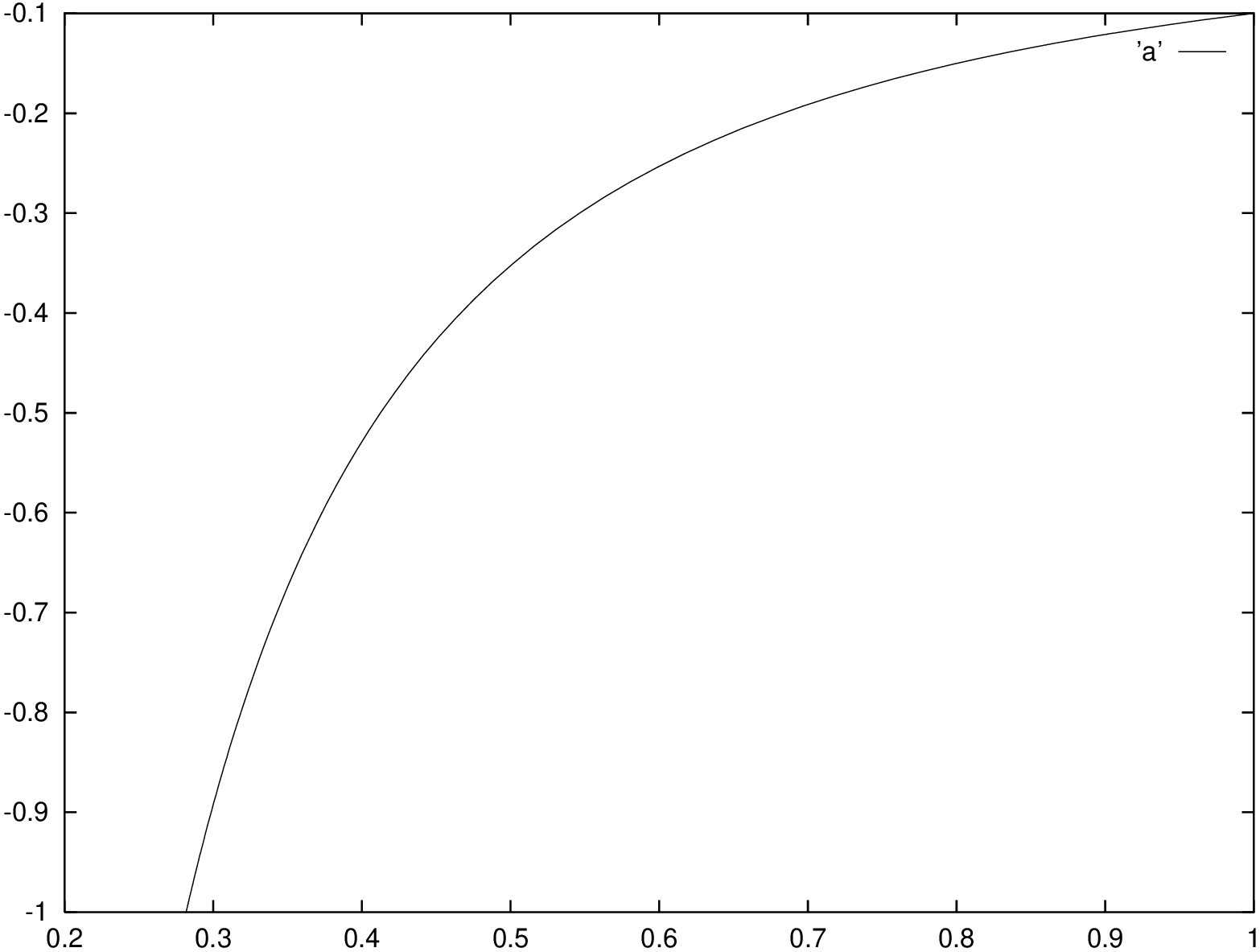}}
\centerline{$g_2$}
\end{minipage}
\hfil
\begin{minipage}{10cm}
\begin{minipage}{4.5cm}
	\epsfxsize=4cm
	\epsfysize=4cm
	\centerline{\epsffile{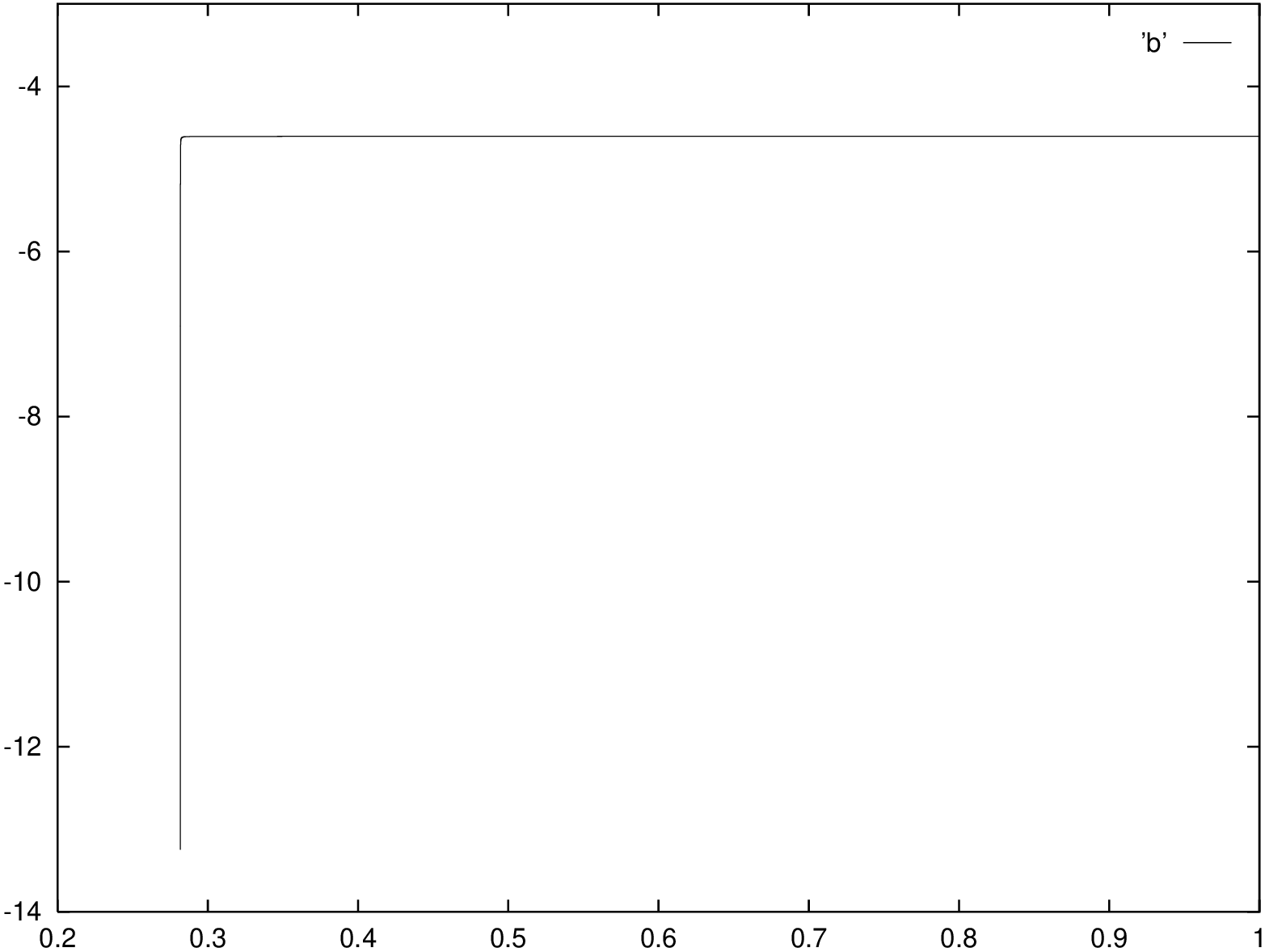}}
\centerline{$g_4$}
\end{minipage}
\hfill
\begin{minipage}{4.5cm}
	\epsfxsize=4cm
	\epsfysize=4cm
	\centerline{\epsffile{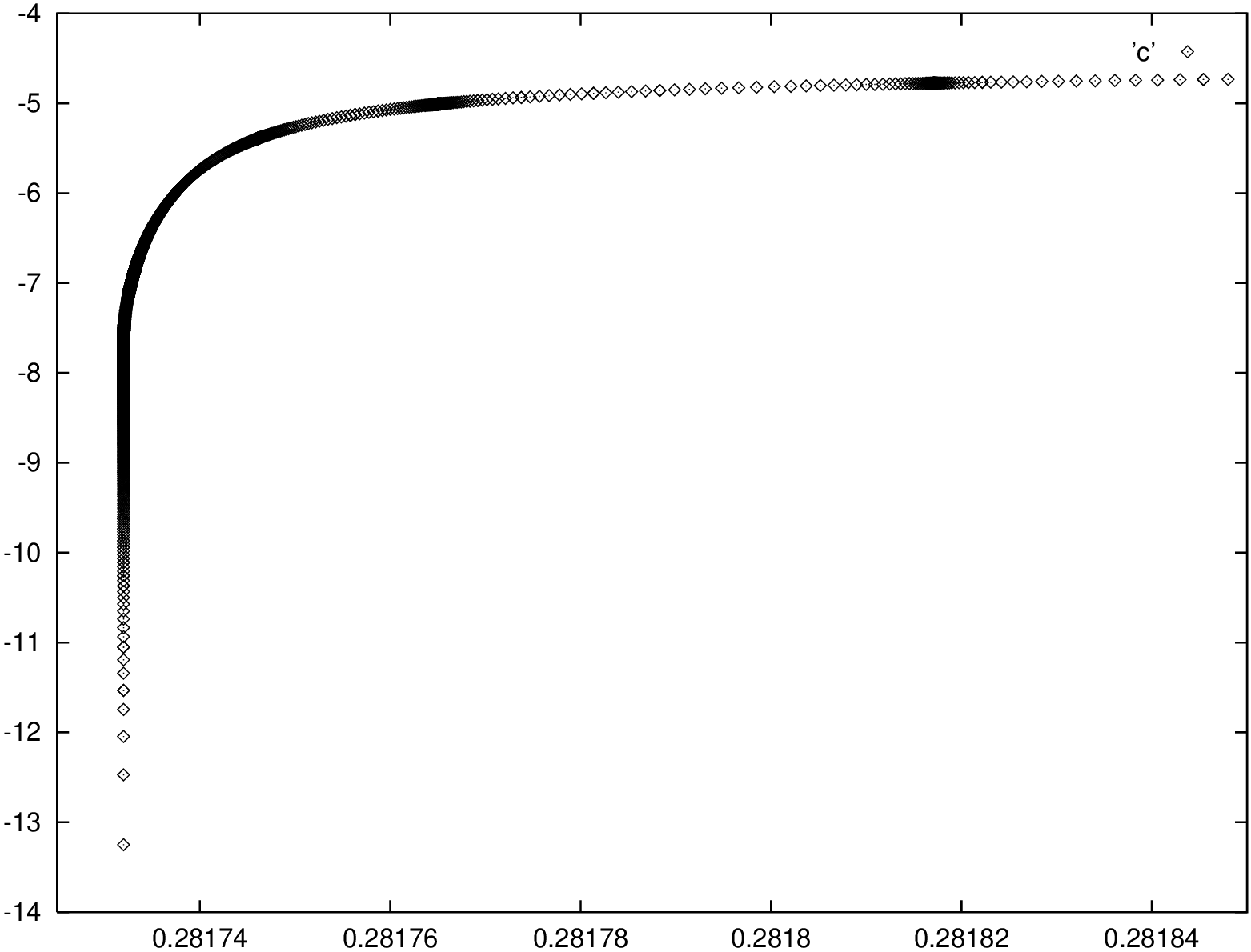}}
\centerline{$g_4$}
\end{minipage}
\end{minipage}
\begin{minipage}{10cm}
\begin{minipage}{4.5cm}
	\epsfxsize=4cm
	\epsfysize=4cm
	\centerline{\epsffile{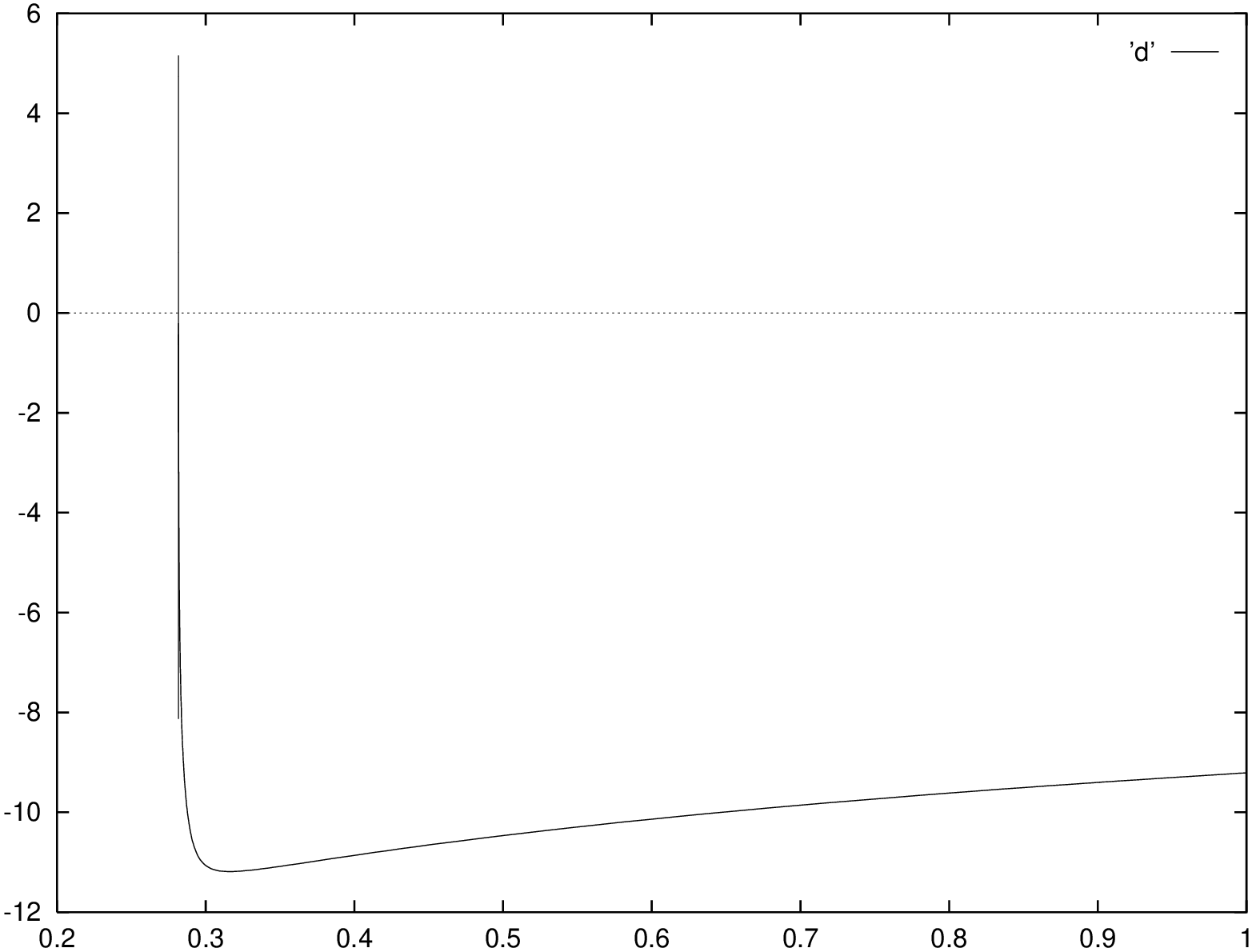}}
\centerline{$g_6$}
\end{minipage}
\hfill
\begin{minipage}{4.5cm}
	\epsfxsize=4cm
	\epsfysize=4cm
	\centerline{\epsffile{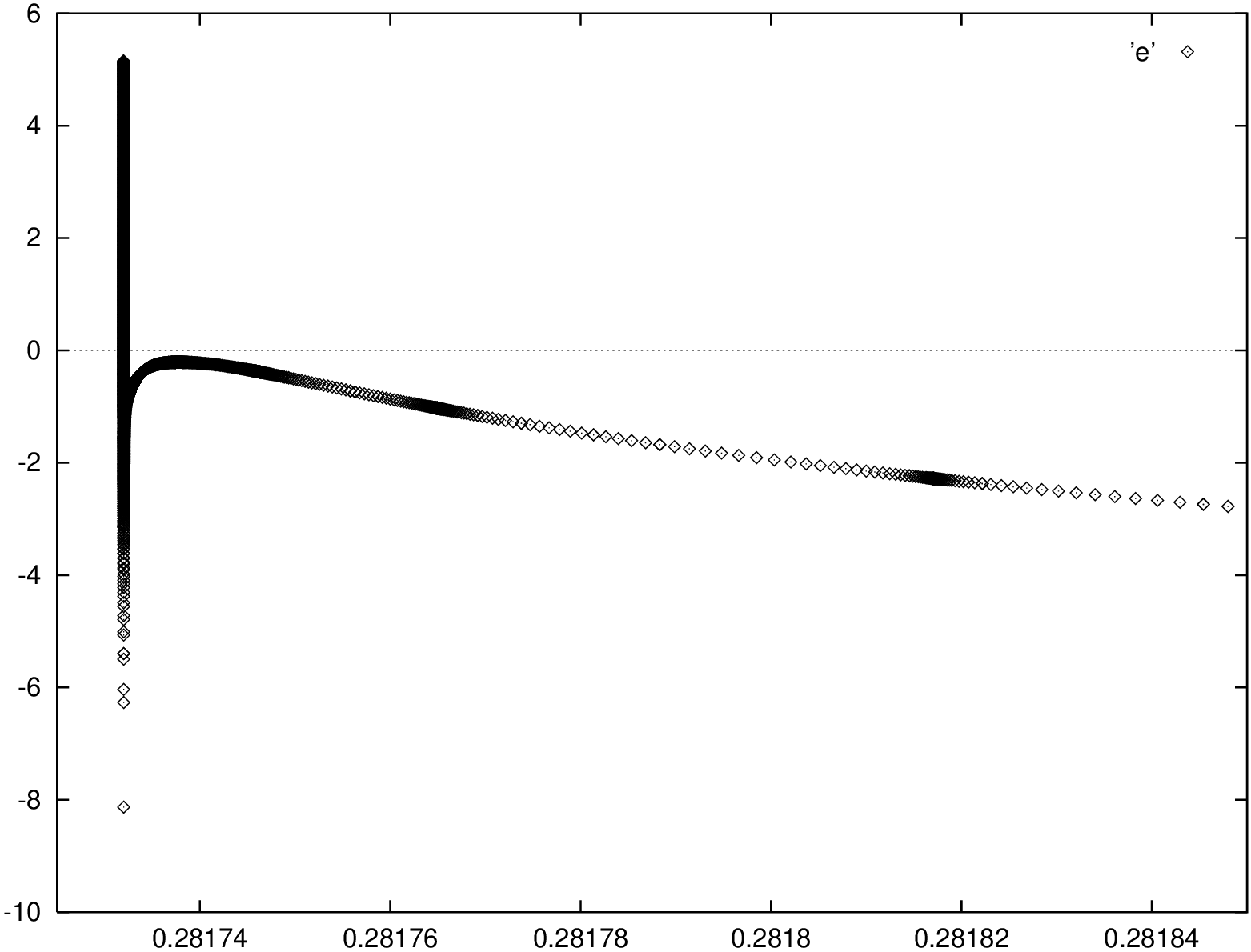}}
\centerline{$g_6$}
\end{minipage}
\end{minipage}
\begin{minipage}{10cm}
\begin{minipage}{4.5cm}
	\epsfxsize=4cm
	\epsfysize=4cm
	\centerline{\epsffile{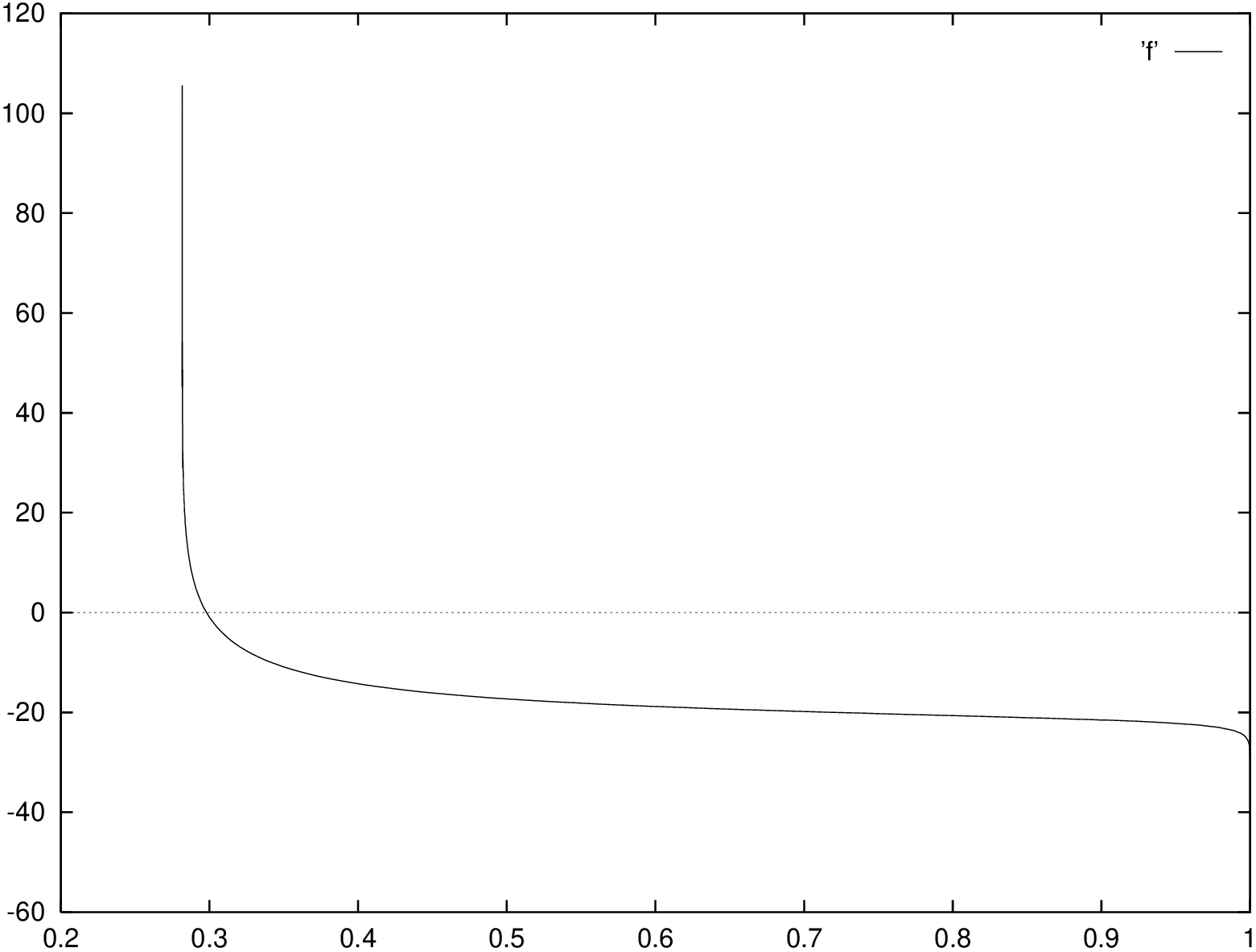}}
\centerline{$g_{20}$}
\end{minipage}
\hfill
\begin{minipage}{4.5cm}
	\epsfxsize=4cm
	\epsfysize=4cm
	\centerline{\epsffile{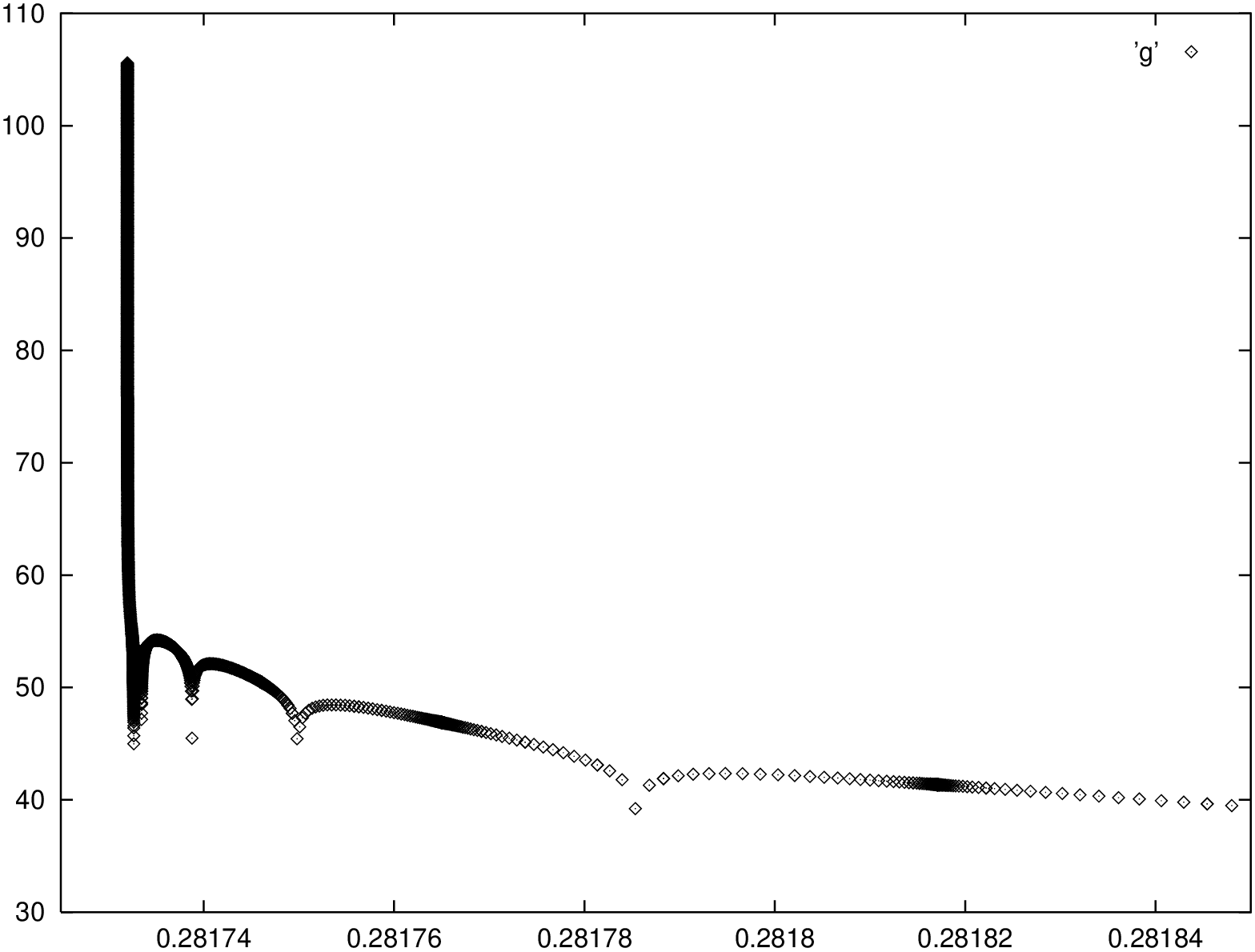}}
\centerline{$g_{20}$}
\end{minipage}
\end{minipage}
\end{figure}
\begin{figure}
\begin{minipage}{10cm}
\begin{minipage}{4.5cm}
	\epsfxsize=4cm
	\epsfysize=4cm
	\centerline{\epsffile{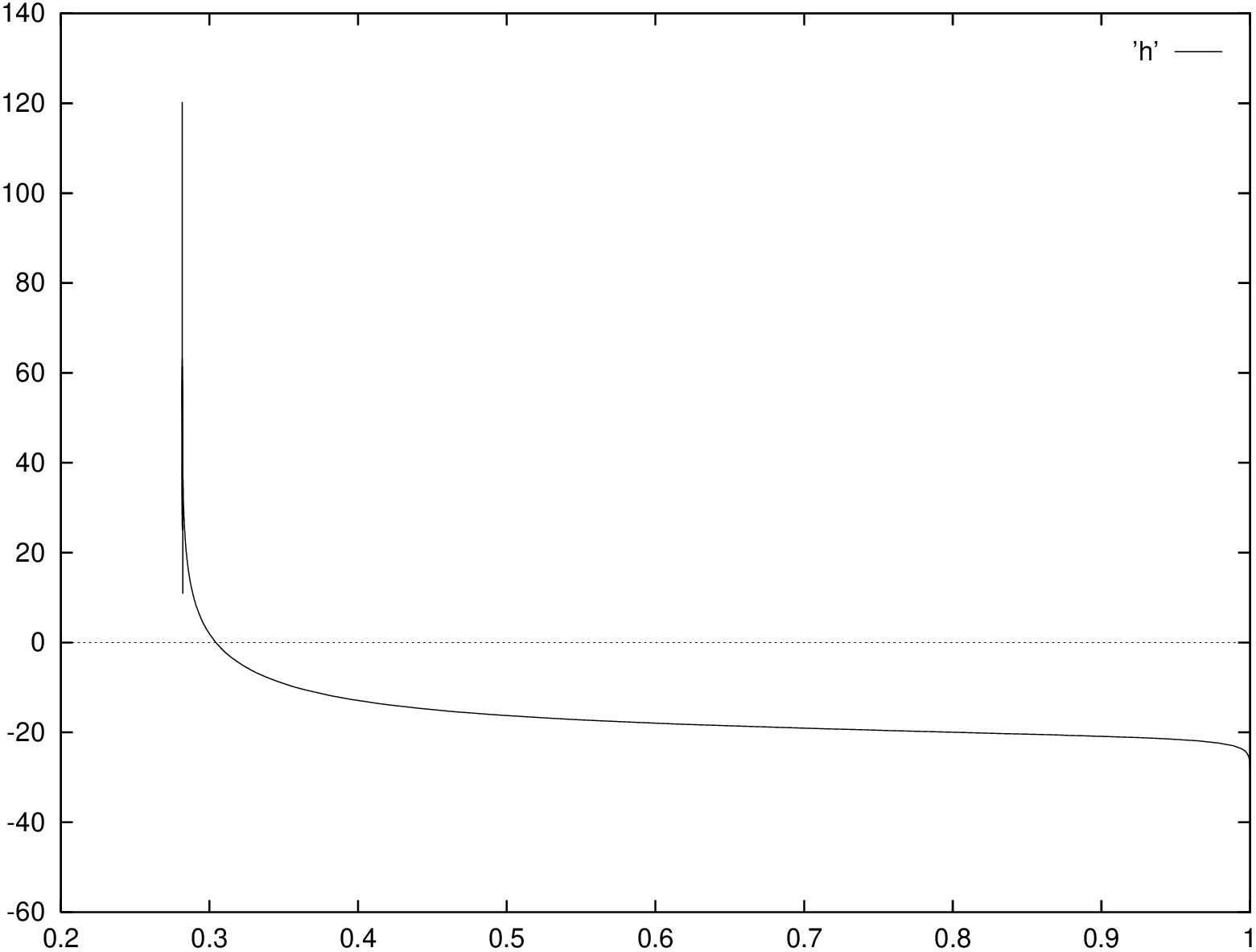}}
\centerline{$g_{22}$}
\end{minipage}
\hfill
\begin{minipage}{4.5cm}
	\epsfxsize=4cm
	\epsfysize=4cm
	\centerline{\epsffile{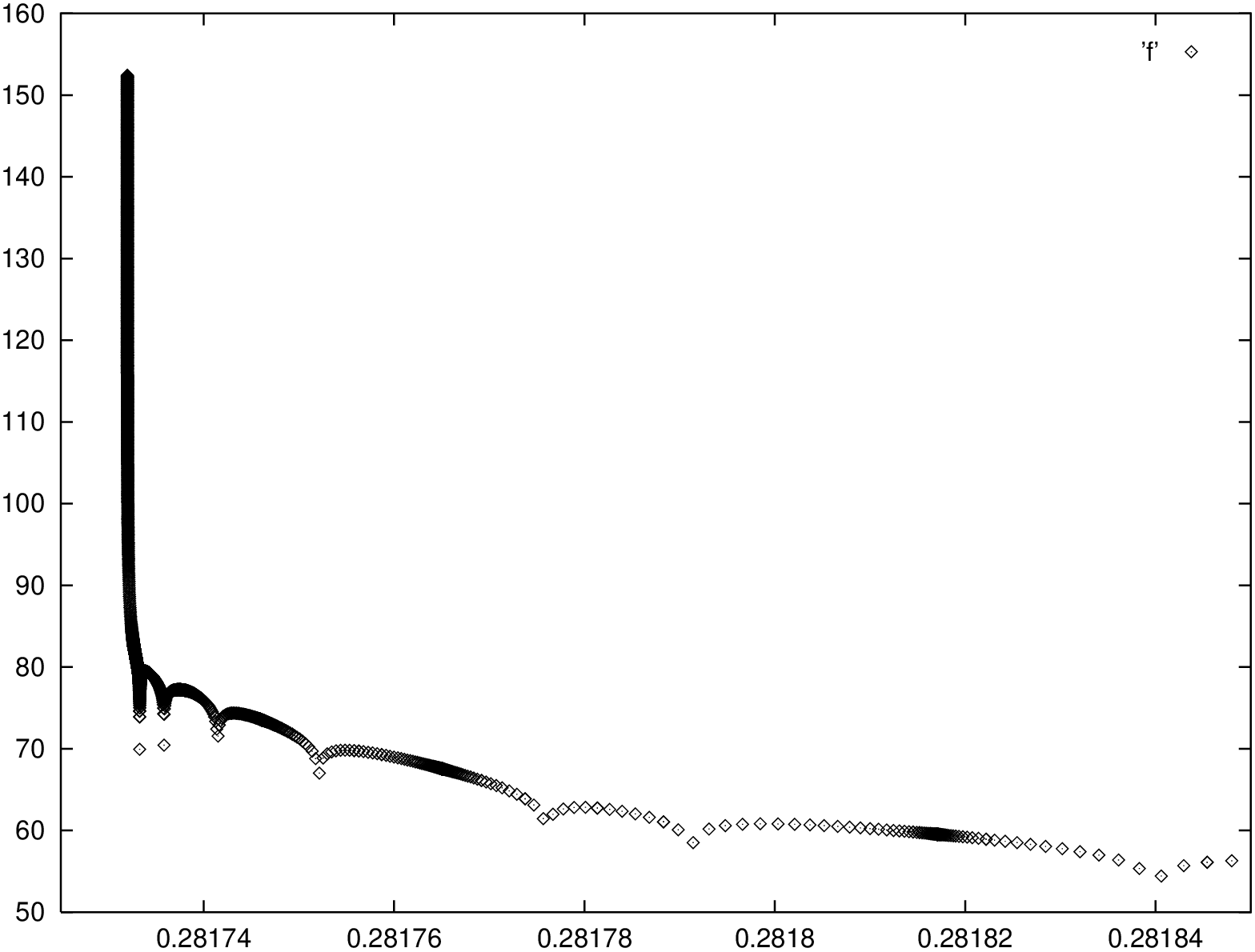}}
\centerline{$g_{22}$}
\end{minipage}
\end{minipage}
\caption{The evolution of $\log|g_n|$ for some values of $n$ at $N=11$, with
the initial conditions $g_2(\Lambda)=-0.1\Lambda^2$, $g_4(\Lambda)=0.01$,
$g_6(\Lambda)=10^{-4}\Lambda^{-2}$ and $g_j(\Lambda)=0.0$ for 
$j=4,\cdots,11$. Where $g_n(k)$ changes sign during the oscillations 
for $n>4$ there $\log|g_n(k)|$ shows a cusp whose altitude is finite due 
to the finite resolution of the $k$ values on the plot.  The points shown 
in the more detailed curves are separated by $10^3$ iterations.}
\end{figure}

It is interesting to observe the derivative of few beta functions with respect to the initial
value of a nonrenormalizable coupling constant, $g_6(\Lambda)$,
\be
{\partial\beta_n(k)\over\partial g_6(\Lambda)}
\label{derbfv}
\ee
what is presented in fig. 3. The derivatives are small in the ultraviolet
scaling regime and diverge at the instability. This divergence shows the presence of a strong 
amplification mechanism due to the instability what lets the small details of the microscopic 
interaction felt in the vicinity of $k=k_{cr}(0)$. By assuming the singular
behavior $\tilde k^{-\nu_n}$ for (\ref{derbfv}) as far as the $k$ dependence is
concerned at $k\approx k_{cr}(0)$ and the perturbative $O(\Lambda^2)$ suppression mechanism 
in the ultraviolet scaling regime for $g_6$, one arrives at
\be
{\partial\beta_n(k)\over\partial g_6(\Lambda)}\approx\tilde k^{-\nu_n}\Lambda^{-2},
\ee
what reveals a sensitivity of the dynamics at the scale $k$ to the values of the
nonrenormalizable parameters at 
\be
\Lambda\approx\tilde k^{-\nu_n/2}.
\ee

\begin{figure}\label{fbfct}
\begin{minipage}{10cm}
\begin{minipage}{4.5cm}
	\epsfxsize=4cm
	\epsfysize=5cm
	\centerline{\epsffile{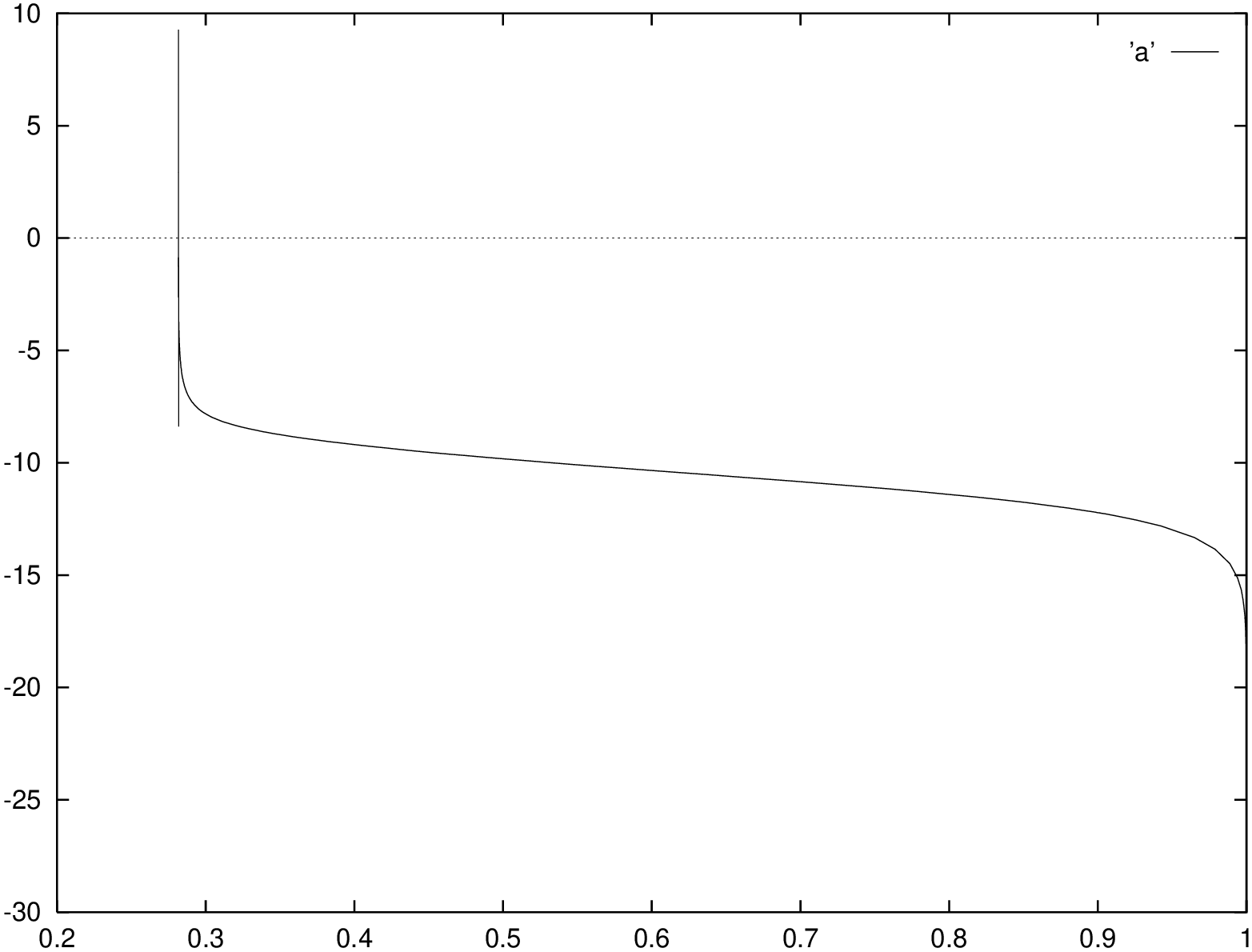}}
\centerline{$\beta_2$}
\end{minipage}
\hfill
\begin{minipage}{4.5cm}
	\epsfxsize=4cm
	\epsfysize=5cm
	\centerline{\epsffile{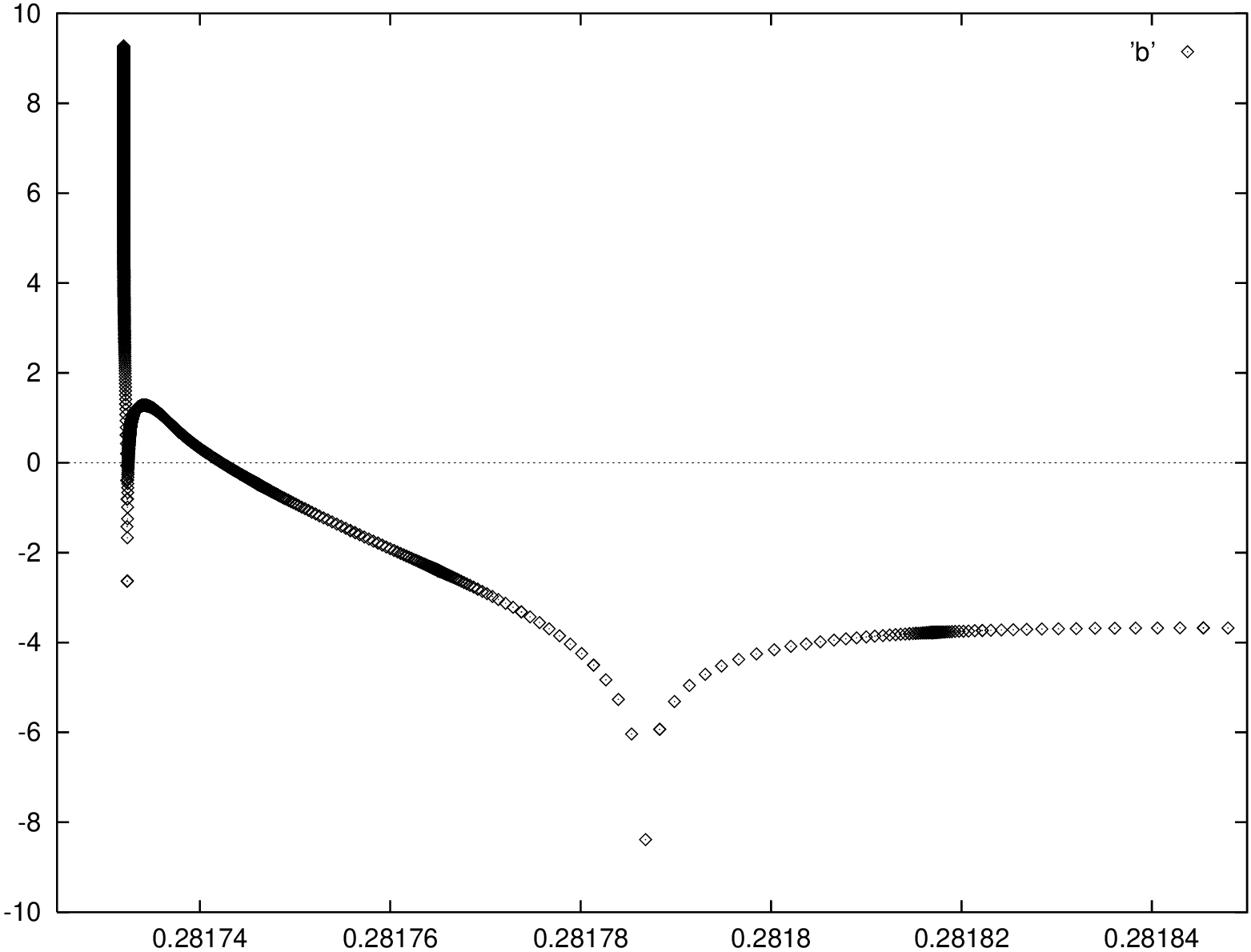}}
\centerline{$\beta_2$}
\end{minipage}
\end{minipage}
\begin{minipage}{10cm}
\begin{minipage}{4.5cm}
	\epsfxsize=4cm
	\epsfysize=5cm
	\centerline{\epsffile{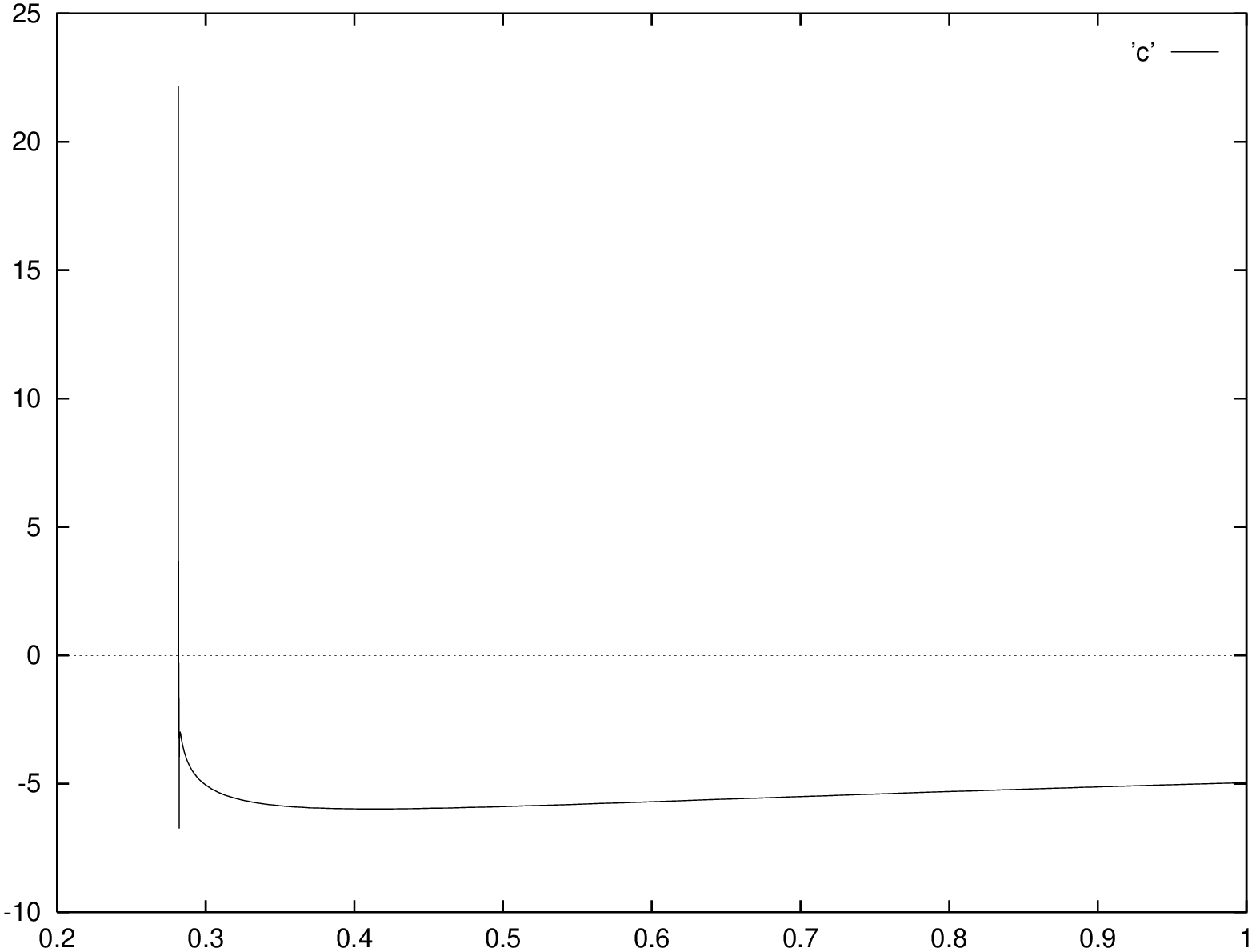}}
\centerline{$\beta_4$}
\end{minipage}
\hfill
\begin{minipage}{4.5cm}
	\epsfxsize=4cm
	\epsfysize=5cm
	\centerline{\epsffile{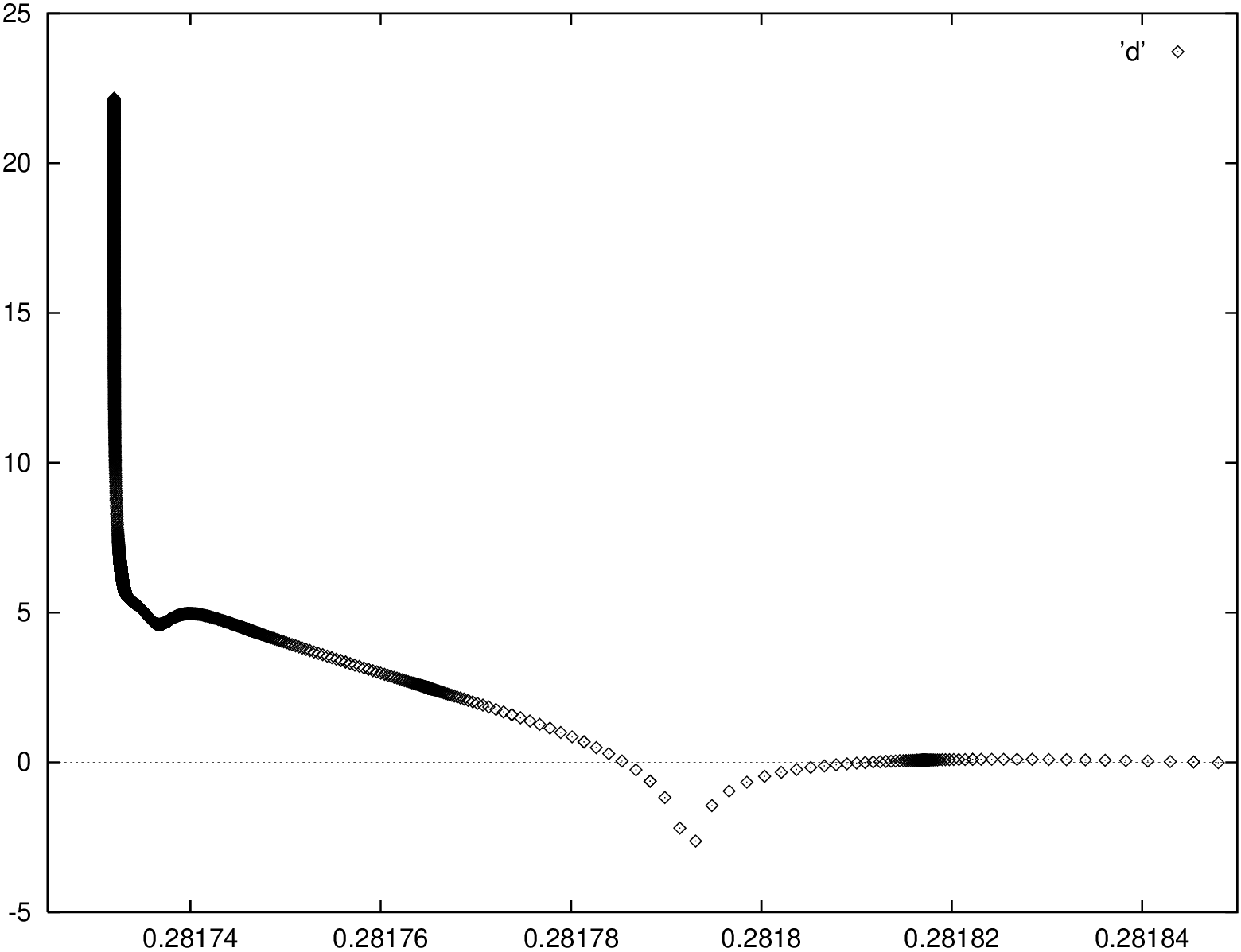}}
\centerline{$\beta_4$}
\end{minipage}
\end{minipage}
\begin{minipage}{10cm}
\begin{minipage}{4.5cm}
	\epsfxsize=4cm
	\epsfysize=5cm
	\centerline{\epsffile{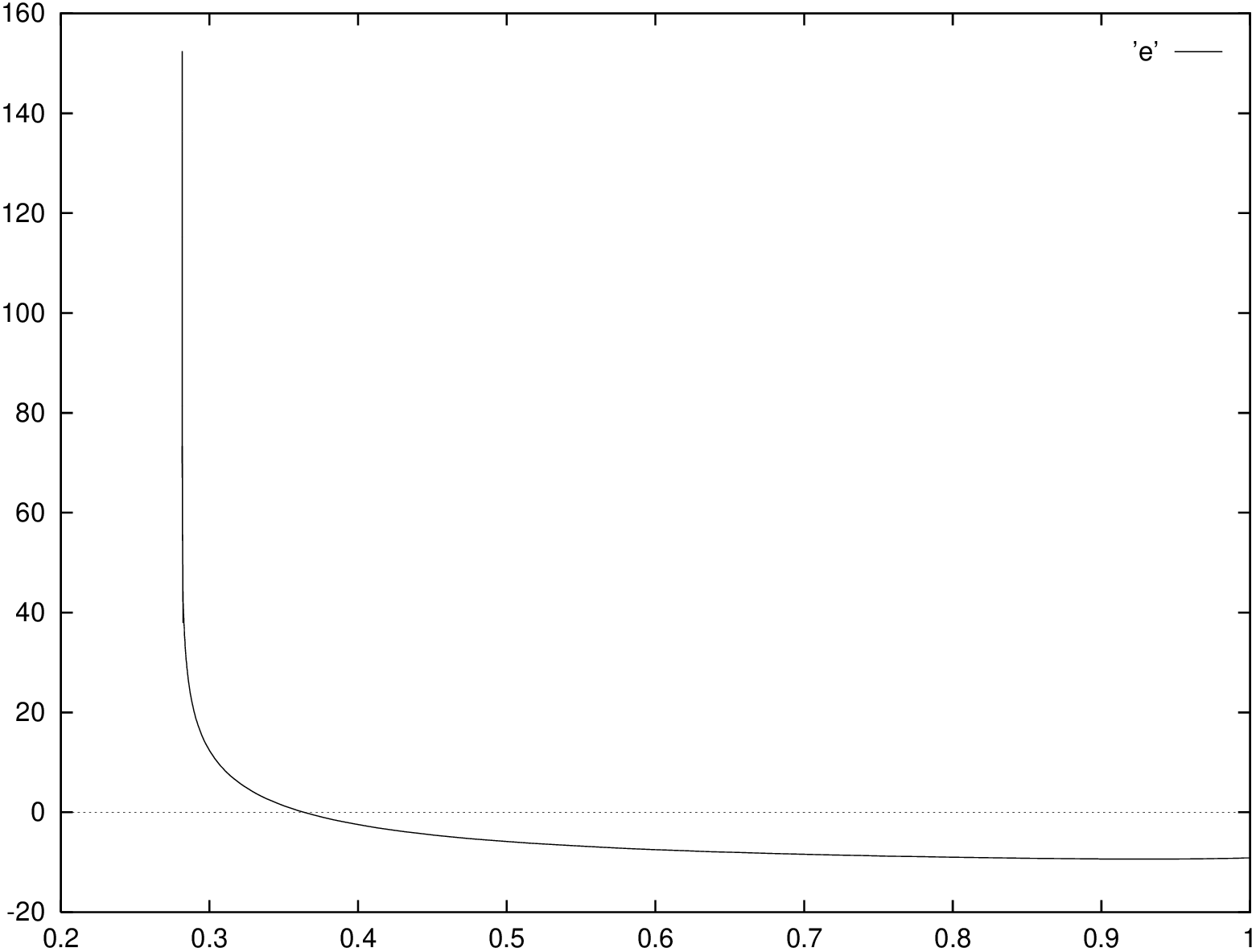}}
\centerline{$\beta_{22}$}
\end{minipage}
\hfill
\begin{minipage}{4.5cm}
	\epsfxsize=4cm
	\epsfysize=5cm
	\centerline{\epsffile{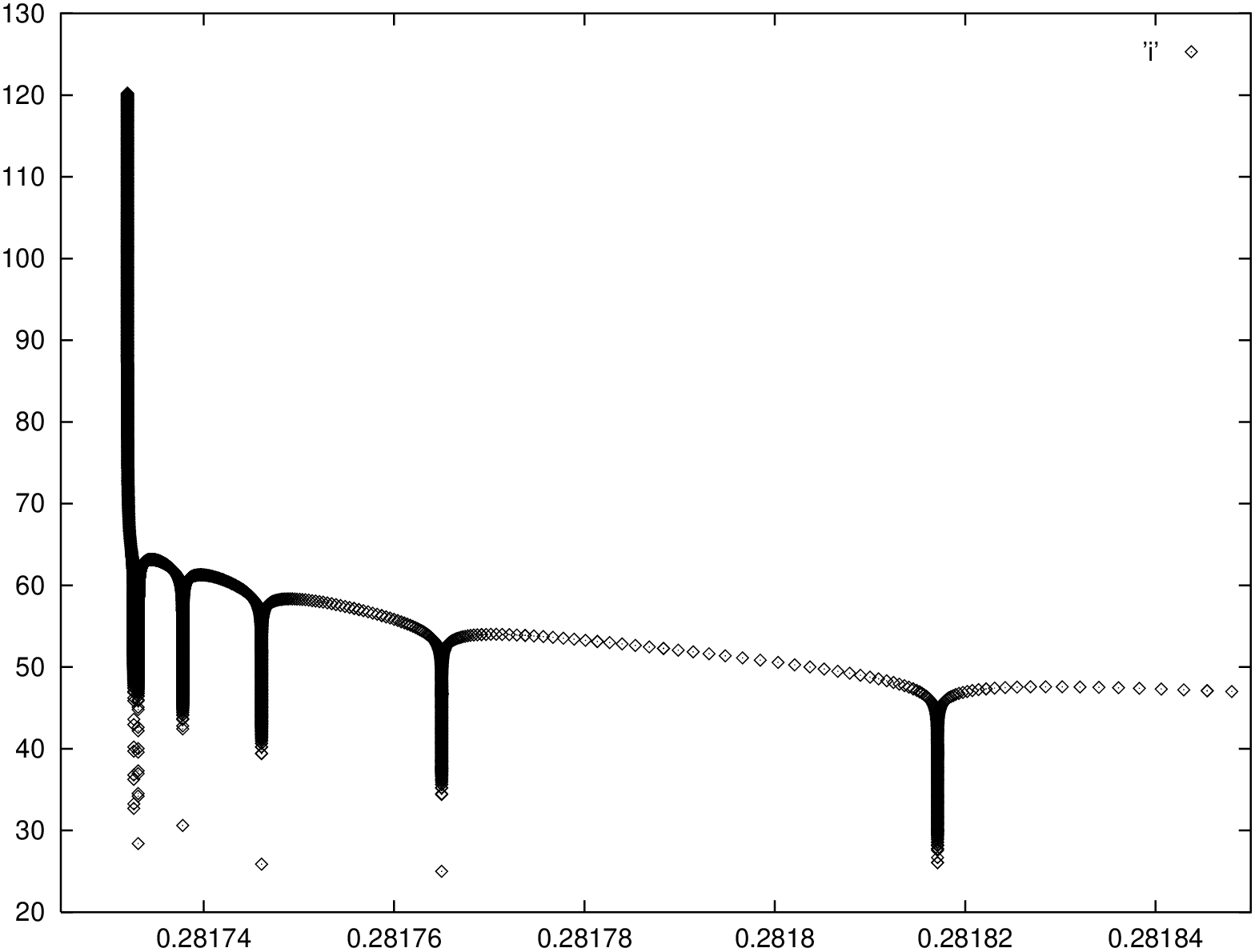}}
\centerline{$\beta_{22}$}
\end{minipage}
\end{minipage}
\caption{The evolution of $\log|{\partial\beta_n(k)\over\partial g_6(\Lambda)}|$
with $N=11$.}
\end{figure}

We note that there is no difficulty to extend the analysis by including
the complete $O(\partial^2)$ order in the gradient expansion. The preliminary
numerical results show no qualitative modification of the behavior presented
above. The divergence of the coupling constants defined by expanding $U_k(\Phi)$
and $Z_k(\Phi)$ increases with the order of $\Phi$ what suggests that the relevant 
operator of this scaling regime might be nonlocal.

We close this Section by few remarks about the interior, unstable region.
The problem in this region is related to the appearance of the saddle
points at the blocking transformation what correspond to plane waves 
and make the renormalization group step very involved.
Apart from the technical complications due to the inhomogeneous background
field a conceptual problem arises, namely the renormalization at the
tree level. This happens when there is a saddle point with nonvanishing
length scale\cite{herve} what gives an $O(\hbar^0)$ contribution to
the blocking relations. The novel feature of the tree level renormalization is
that it does not fit into the usual classification scheme of the coupling 
constants what is based on the inspection of the loop integrals. The classical
differential equations may produce richer and more singular dependence in the
coupling constants than the polynomes of the perturbation expansion. This actually
happens in the unstable region and the resulting blocking relation offers no
guarantee that change in the blocked action will be small, infinitesimal if the 
the cutoff is decreased in a small, infinitesimal amount.
By {\em assuming} that the renormalized trajectory, i.e. the path integral 
possesses finite derivative with respect the cutoff within the unstable region
we have an additional consistency relations. It simplifies the problem and the 
important saddle point contributions to the blocking 
can be resummed with the result\cite{jeani}
\be
U_k(\Phi)=-\hf k^2\Phi^2+c(k),\label{inter}
\ee
where $c(k)$ is chosen to have continuous potential at the singular curve
$k^2=k^2_{cr}(\Phi)$. Observe that this potential is indeed a "fixed point", i.e.
stays invariant under blocking. For each plane wave saddle point there is a
zero mode what is related to the translation in the direction of the wave vector.
The integration over the zero modes finally restores the translation invariance.

Another, independent support of this form comes from a simple exact relation 
given in terms of the variables in their original, dimensional form\cite{jeani},
\be
{\partial V_k(\Phi)\over\partial k}{\partial W_k(\Phi)\over\partial\Phi}=
{\partial W_k(\Phi)\over\partial k}{\partial V_k(\Phi)\over\partial\Phi},
\ee
where
\bea
V_k(\Phi)&=&k^{1-d}{\partial U_k(\Phi)\over\partial k},\nonu
W_k(\Phi)&=&k^2+U_k''(\Phi).
\eea
It was derived by using the gradient expansion only, without making reference 
to the loop expansion. This is a higher order differential equation which has
spurious solutions but it is at least satisfied by (\ref{inter}).
We believe that the simplicity of our result, (\ref{inter}), originates from the 
same balance between the energy and entropy as in the case of the Maxwell construction
except that the role of the domains is played by the zero modes of each saddle
point plane wave with $k<k_{cr}(0)$.

\section{Finite Temperature}
The vacuum of the symmetry broken theory is in the stable region so the
singularities of the coupling constants studied in this work are not 
obviously important for the fluctuations around the vacuum. But the
singularity reappear when the symmetry is broken by controlling external
environment variables, such as the temperature. Imagine the cooling of
a ferromagnet slightly above the Curie point or the hot Universe before the spontaneous
breakdown of a global symmetry group. The order parameter is at its symmetrical
value $\Phi=0$ in the high temperature phase and the fluctuations are characterized 
by the coupling constants at $\Phi=0$ what diverge and reflect the nonuniversal
behavior investigated above. When the system arrives at the symmetry 
broken phase then the order parameter remains close to the minimum of the effective 
potential what is outside of the spinodial unstable region so long as the
time evolution is adiabatic. When $T<<T_{cr}$ the
vacuum is far from the unstable region what is detected by the large 
amplitude fluctuations only. Thus the saddle points of the inner, unstable region 
influence the evolution around $T\approx T_{cr}$, new scaling law can be observed
in the vicinity of the phase transition.

To outline the procedure at finite temperature we note
that the blocking should be made in three space only in order to preserve the
value of the physical temperature on the renormalized trajectory. Such an
anisotrop blocking in the Euclidean space-time where the cutoff $k$ refers to the
three space and the fluctuations in time are eliminated at the one loop level
yields the equation\cite{sbft}, 
\bea
k{\partial\over\partial k}U_k(\Phi)&=&-{T\Omega_3k^3\over2(2\pi)^3}
\sum_{n=-\infty}^\infty\log\left[\omega^2_n+k^2+U_k''(\Phi)\right]\\
&=&-{T\Omega_3k^3\over2(2\pi)^3}\left\{{1\over T}\sqrt{k^2+U_k''(\Phi)}
+2\log\left[1-e^{-{1\over T}\sqrt{k^2+U_k''(\Phi)}}\right]\right\},\nonumber
\eea
where $\omega_n=2\pi nT$. Another simpler strategy is to eliminate the
$|n|>n_0$ Matsubara modes perturbatively and to apply the blocking
for $|n|\le n_0$ only\cite{ppp}. The beta functions derived by
either method diverge in the vacuum at $\Phi=k=0$ as $T\to T_{cr}$ from above 
and one expects a singular structure to appear what is similar to what was
presented above in four dimensions at $k=k_{cr}(0)$.

\section{Summary}
We presented evidences that radically different scaling laws are present in
the $\phi^4$ model with spontaneous symmetry breaking at high and low energies. 
The system appears nonuniversal at the phase transition where the influence of the
nonrenormalizable coupling constants what is suppressed at the ultraviolet 
scaling regime can be compensated for by the singular tree level structure of the
condensate formation. It was mentioned that the renormalization group method
can successfully be applied to the spinodial instability, the mixed phase.
The saddle points of the blocking procedure which are plane waves can be
taken into account and the integration over the zero modes restores the
homogeneity of the mixed phase, what is the reminiscent of the Maxwell 
construction. 

It is worthwhile noting that as one increases the 
number of terms retained in the potential then the qualitative behavior
changes completely when the potential was retained up to $O(\Phi^{22})$. We 
believe that the divergences occurring in the gradient expansion suggests the 
presence of nonlocal relevant operators at the low energy scaling regime.

There are several questions left open by these results. What we find the most 
pressing is the classification of the possible nonrenormalizable parameters what 
influences the seemingly nonuniversal dynamics of the phase transition.
A related issue is the more useful application of the amplification mechanism of the 
divergences generated by the instability as a "renormalization group microscope" in 
the coupling constant space to discover the microscopic parameters from the long
distance observables.

%\section{Acknowledgment}

\end{document}